\documentclass[sn-mathphys-num]{sn-jnl}% Math and Physical Sciences Numbered Reference Style 
\usepackage[abs]{overpic}
\usepackage{graphicx}%
\usepackage{multirow}%
\usepackage{amsmath,amssymb,amsfonts}%
\usepackage{amsthm}%
\usepackage{mathrsfs}%
\usepackage[title]{appendix}%
\usepackage{xcolor}%
\usepackage{textcomp}%
\usepackage{manyfoot}%
\usepackage{booktabs}%
\usepackage{algorithm}%
\usepackage{algorithmicx}%
\usepackage{algpseudocode}%
\usepackage{listings}%
\usepackage{scalerel}
\usepackage{tikz}
\usetikzlibrary{svg.path}
\usepackage[abs]{overpic}
\usepackage{hyperref}
\usepackage{graphicx}%
\usepackage{multirow}%
\usepackage{amsmath,amssymb,amsfonts}%
\usepackage{amsthm}%
\usepackage{mathrsfs}%
\usepackage[title]{appendix}%
\usepackage{xcolor}%
\usepackage{textcomp}%
\usepackage{manyfoot}%
\usepackage{booktabs}%
\usepackage{algorithm}%
\usepackage{algorithmicx}%
\usepackage{algpseudocode}%
\usepackage{listings}%
\usepackage{scalerel}
\usepackage{tikz}
\usetikzlibrary{svg.path}
\usepackage{natbib}
\usepackage{algorithm}
\usepackage{algpseudocode}
\usepackage{bm}
\usepackage{overpic}
\usepackage{xcolor}
\usepackage{tabularx}
\usepackage{pict2e}
\usepackage{tikz}
\usepackage{svg}
\usepackage{bm}
\usepackage{hyperref}

\usepackage{pgfplots}
\pgfplotsset{compat=1.18}
\usepackage{comment}
\usepackage{placeins}
\usepackage{algorithm}
\usepackage{algorithmicx}
\usepackage{algpseudocode}

\definecolor{orcidlogocol}{HTML}{A6CE39}
\tikzset{
  orcidlogo/.pic={
    \fill[orcidlogocol] svg{M256,128c0,70.7-57.3,128-128,128C57.3,256,0,198.7,0,128C0,57.3,57.3,0,128,0C198.7,0,256,57.3,256,128z};
    \fill[white] svg{M86.3,186.2H70.9V79.1h15.4v48.4V186.2z}
                 svg{M108.9,79.1h41.6c39.6,0,57,28.3,57,53.6c0,27.5-21.5,53.6-56.8,53.6h-41.8V79.1z M124.3,172.4h24.5c34.9,0,42.9-26.5,42.9-39.7c0-21.5-13.7-39.7-43.7-39.7h-23.7V172.4z}
                 svg{M88.7,56.8c0,5.5-4.5,10.1-10.1,10.1c-5.6,0-10.1-4.6-10.1-10.1c0-5.6,4.5-10.1,10.1-10.1C84.2,46.7,88.7,51.3,88.7,56.8z};
  }
}

\newcommand\orcidicon[1]{\href{https://orcid.org/#1}{\mbox{\scalerel*{
\begin{tikzpicture}[yscale=-1,transform shape]
\pic{orcidlogo};
\end{tikzpicture}
}{|}}}}

\usepackage{hyperref}
\theoremstyle{thmstyleone}%
\theoremstyle{thmstyletwo}%

\theoremstyle{thmstylethree}%

\usepackage{acronym}
\acrodef{PIV}[PIV]{{Particle Image Velocimetry}}
\acrodef{PTV}[PTV]{{Particle Tracking Velocimetry}}
\acrodef{POD}[POD]{{Proper Orthogonal Decomposition}}
\acrodef{STB}[STB]{{Shake-The-Box}}
\acrodef{LPT}[LPT]{{Lagrangian Particle Tracking}}
\acrodef{RBF}[RBF]{Radial Basis Function}
\acrodef{DNS}[DNS]{{Direct Numerical Simulation}}
\acrodef{KNN}[KNN]{{K-Nearest Neighbor}}

\usepackage{longtable}  % For tables that can span multiple pages
\usepackage{amsmath}    % For math symbols
\usepackage{amsfonts}   % For fonts like mathbb

\begin{document}

\title[Article Title]{Meshless Super-Resolution of Scattered Data via constrained \textcolor{black}{Radial Basis Functions} and \textcolor{black}{K-Nearest-Neighbors}-Driven Densification}


\author*[1]{\fnm{Iacopo} \sur{Tirelli}}\email{iacopo.tirelli@uc3m.es \orcidicon{0000-0001-7623-1161}}

\author[2]{\fnm{Miguel Alfonso} \sur{Mendez}}\email{mendez@vki.ac.be \orcidicon{0000-0002-1115-2187}}

\author[1]{\fnm{Andrea} \sur{Ianiro}}\email{aianiro@ing.uc3m.es \orcidicon{0000-0001-7342-4814}}

\author[1]{\fnm{Stefano} \sur{Discetti}}\email{sdiscett@ing.uc3m.es \orcidicon{0000-0001-9025-1505}}

\affil*[1]{\orgdiv{Department of Aerospace Engineering}, \orgname{Universidad Carlos III de Madrid}, \orgaddress{\street{Avda. Universidad 30}, \city{Leganés}, \postcode{28911}, \state{Madrid}, \country{Spain}}}

\affil[2]{\orgdiv{Environmental and Applied Fluid Dynamics}, \orgname{von Karman Institute for Fluid Dynamics}, \orgaddress{\street{Waterloosesteenweg 72}, \city{Sint-Genesius-Rode}, \postcode{1640}, \state{Bruxelles}, \country{Belgium}}}


% General remark: the abstract seems a bit too long to me. It seems more an introduction than an abstract. The part below could be moved to the introduction. 

%These approaches have become ubiquitous in many areas of applied science, with their original form known as Karhunen-Loève transform (KL) in applied mathematics, Empirical Orthogonal Functions (EOF) in meteorology and climatology, and Proper Orthogonal Decomposition (POD) in fluid mechanics. 

% note: the PCA is commonly used also in cases where the grid is not structured. The key, however, is that it must be fixed in space.

\abstract{
	We propose a novel meshless method to achieve super-resolution from scattered data obtained from sparse, randomly-positioned sensors such as the particle tracers of particle tracking velocimetry. The method combines K-Nearest Neighbor Particle Tracking Velocimetry \citep[KNN-PTV,][]{tirelli2023end} with meshless Proper Orthogonal Decomposition  \citep[meshless POD,][]{tirelli2024meshless} and constrained Radial Basis Function regression \citep[c-RBFs,][]{sperotto2022meshless}. The main idea is to use KNN-PTV to enhance the spatial resolution of flow fields by blending data from \textit{locally similar} flow regions available in the time series. This \textit{similarity} is assessed in terms of statistical coherency with leading features, identified by meshless POD directly on the scattered data without the need to first interpolate onto a grid, but instead relying on RBFs to compute all the relevant inner products. Lastly, the proposed approach uses the c-RBF on the denser scattered distributions to derive an analytical representation of the flow fields that incorporates physical constraints. This combination is \textit{meshless} because it does not require the definition of a grid at any step, thus providing flexibility in handling complex geometries. \textcolor{black}{An ablation study on the role of penalties and physical constraints demonstrates their key contribution in regularizing the regression and ensuring physically-consistent reconstructions.} The algorithm is validated on $3$D measurements of a jet flow in air. The assessment covers three key aspects: statistics, spectra and modal analysis. The proposed method is evaluated against standard Particle Image Velocimetry, KNN-PTV, and c-RBFs. The results demonstrate improved accuracy, with an average error on the order of $10\%$, compared to $12-13\%$ for the other methods. When considering reduced-order reconstructions, the error is even halved. Additionally, the proposed method \textcolor{black}{exhibits a higher frequency cut-off (based on reaching a noise floor) than the one} observed in the competing approaches.  \textcolor{black}{A qualitative comparison highlights that the KNN-driven densification of the particle distribution also enhances the quality of velocity derivatives and related quantities.
}}


\maketitle

\section{Introduction}
\label{Sec:Intro}

The measurement of detailed quantitative field data of turbulent flows poses significant challenges due to the vast range of spatial and temporal scales involved, which widens with increasing Reynolds numbers. While \acl{PIV} \citep[\acs{PIV},][]{raffel2018particle} has become a robust tool for this purpose \citep{westerweel2013particle}, its ability to resolve turbulent scales is constrained by both the sensor size and the inter-particle spacing in the images. \acused{PIV}

Although spatial resolution can potentially be improved by exploiting temporal consistency in short sequences \citep{sciacchitano2012multi,schneiders2014time}, it seems that two-frame \ac{PIV} has reached its physical limits in terms of dynamic spatial range. \citet{kahler2012resolution} demonstrated that, for average flow fields, the particle diameter imposes a key limitation on cross-correlation-based methods, confirming particle tracking as the most suitable approach to enhance spatial resolution. On instantaneous measurements, dense vector fields can be obtained with super-resolution \acl{PTV} \citep[super-resolution \acs{PTV},][]{keane1995super}, although its reliability is more dependent on image quality than standard cross-correlation. \acused{PTV} In this context, the term super-resolution refers to the integration of PTV within PIV interrogation windows, enabling velocity measurements that exceed the spatial resolution limits of traditional cross-correlation methods.

In super-resolution PTV, the spatial resolution is determined by the smaller mean spacing between particles and their displacement between light pulses. In $3$D velocimetry, the superiority of particle tracking over cross-correlation is even more established than that of the planar counterpart. The lower computational cost of evaluation, the lower risk of ambiguity in particle pairing, and the use of multiple cameras that allow disambiguating particles overlapping in an image using other views are factors that play in favour of particle tracking.

It is nevertheless common practice to interpolate the data on a Eulerian grid for postprocessing and visualization. This process inevitably introduces unnecessary filtering effects and artifacts due to interpolation \citep{scarano2003theory}, affecting the final accuracy of the measured flow fields. 

\begin{figure}[t]
	\centering
	\begin{overpic}[width=0.95\textwidth, unit = 1mm]
		{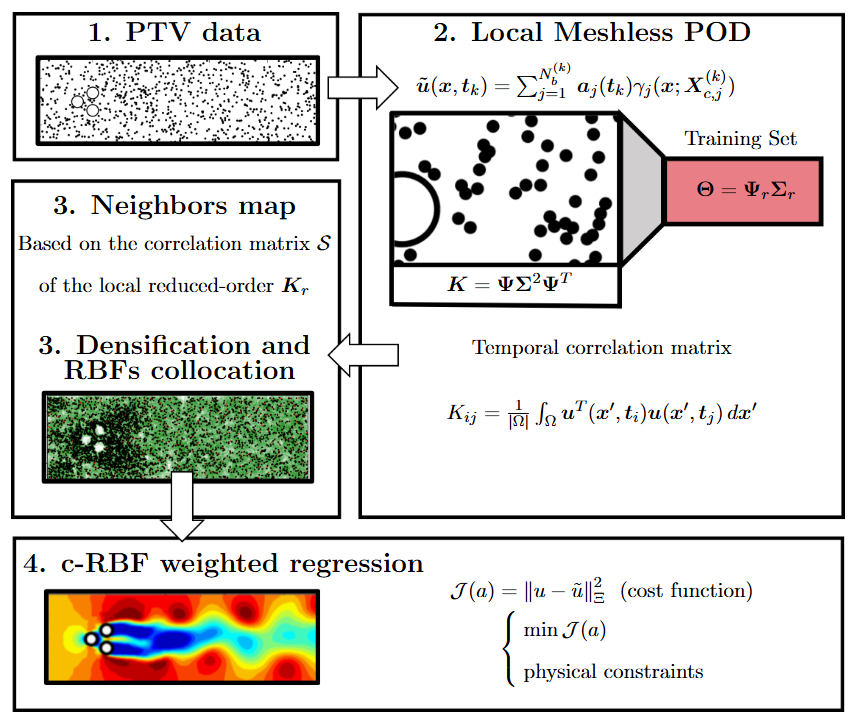}
		
	\end{overpic}
	\caption{Flowchart of the proposed algorithm. Step 1: extraction of information from particles (PTV or LPT); step 2: local meshless POD directly on the particles within the subdomains into which the domain is partitioned; step 3: computing optimal number of neighbours and then increasing particle density; step 4: weighted regression trough c-RBFs to achieve analytical high-resolution approximation of the velocity field. }
	\label{Fig:flowchart}
\end{figure}

In this work, we adopt and extend the meshless super-resolution framework introduced by \citet{sperotto2022meshless,ratz2024meshless}, which seek to eliminate interpolation and grids from all post-processing steps in tracking velocimetry. In this framework, the notion of super-resolution refers to the derivation of a mesh-independent analytical representation of the velocity field, allowing predictions at any point while simultaneously enforcing physical constraints (e.g., no-slip condition, mass conservation, etc.). 
This formalism is based on constrained Radial Basis Functions (c-RBFs) \acused{RBF}regression \citep{Mendez_BOOK}, placed in the measurement domain and shaped in ways that account for seeding density inhomogeneities. Nevertheless, when processing instantaneous fields, the accuracy and feasibility of the method are constrained by particle availability. Regions with sparse sampling require larger \acp{RBF}, which can ultimately lead to significant low-pass filtering of the velocity field. The method proposed in this work offers a fully meshless approach to increase seeding density prior to a c-\acp{RBF} regression.

A large class of methods leverages statistical evidence of correlation to increase the particle density in instantaneous realizations by ``borrowing'' information from other snapshots of the dataset. The underpinning of these methods stands upon recognizing that each velocity vector realization represents a sample from an underlying statistical distribution. This principle gave rise to established techniques for resolution enhancement of turbulence statistics, see e.g single-pixel correlation \citep{meinhart2000piv,westerweel2004single, scharnowski2012reynolds} and ensemble PTV \citep{cowen1997hybrid,kahler2012resolution, aguera2016ensemble}. Recent advances in data-driven and machine-learning algorithms pushed this concept to individual samples \citep{discetti2022machine}. A first effort in this direction was followed by \citet{cortina2021sparse}, who proposed a Data-Enhanced PTV (DEPTV). The main idea was that high-resolution \ac{POD} modes can be obtained by progressively filling linear-stochastic estimates based on projection on a temporal basis obtained by standard \ac{PIV} data analysis. \textcolor{black}{While this concept is well established for time-averaged quantities, similar strategies have also been successfully applied to the reconstruction of instantaneous flow fields. Notably, approaches based on Linear Stochastic Estimation and \ac{POD} \citep{bonnet1994stochastic,durgesh2010multi} and Extended \ac{POD} \citep{boree2003extended} provide evidence that such methodologies are not limited to averaged quantities.}
The DEPTV assumes a linear mapping between low- and high-resolution basis. This hypothesis was progressively weakened by switching to locally-linear methods and fully nonlinear methods. In the first category, K-Nearest Neighbor Particle Tracking Velocimetry \citep[KNN-PTV,][]{tirelli2023end} stands out for its simplicity of implementation. KNN-PTV is based on the idea of blending particles from different snapshots when local regions are sufficiently similar. The similarity was assessed based on \ac{POD} of the flow fields obtained by standard \ac{PIV} analysis. Thus, intrinsically, the method relies on an Eulerian grid for its implementation. Nonlinear methods can provide better performances but have similar requirements, especially when using deep learning techniques such as Generative Adversarial Networks \citep{deng2019super, guemes2022super} or estimators based on optical flow \citep{cai2019dense,lagemann2021deep, yu2021lightpivnet}. \textcolor{black}{In this context, Graph Neural Networks \citep[GNNs,][]{scarselli2008graph} are particularly suitable due to their ability to perform convolution operations on unstructured data, as detailed by \citet{wu2020comprehensive}, while struggling to handle spatially moving sensors. Consequently, these approaches are restricted to fixed sensor configurations matching those used during training. To overcome these limitations, \citet{fukami2021global} proposed a method that incorporates sparse sensor data into a convolutional neural network (CNN) by approximating local information onto a structured representation while preserving the spatial arrangement of the sensors. This is achieved with a Voronoi tessellation of the unstructured dataset and introducing a mask field that encodes the sensor locations as part of the input. Another emerging class of models in computer vision and scientific machine learning are neural-implicit fields \citep{mildenhall2021nerf}, which represent physical variables as continuous functions of spatial and temporal coordinates through coordinate neural networks. Operating in an unsupervised manner, these models enable significant data compression of spatio-temporal fields by eliminating the need for discrete grids or predefined basis functions. When combined with optical flow techniques, they give rise to the so-called neural optical flow (NOF) framework. Recently, \citet{masker2024neural} proposed a new NOF algorithm that improves on existing methods and allows incorporating physical constraints through penalty terms, hence turning the NOF networks into physics-informed neural network \citep[PINN,][]{raissi2019physics}.  } 

However, the existing methods suffer one or more of the issues in the following list: (1) difficulty in including physical constraints in the formulation; (2) restriction in the mapping capability when imposing linearity;  and (3) rigidity on the choice of the grid, which cannot adapt to the scattering of the data.

In this work, we propose combining KNN-PTV with constrained regression using c-\acp{RBF}. On one hand, c-RBFs eliminate the need for a predefined grid in KNN-PTV. On the other hand, KNN-PTV artificially increases particle density in individual snapshots, generating denser scattered distributions that enable c-\acp{RBF} to produce physically-constrained and highly-accurate super-resolution fields. The key novelty to achieve a fully-meshless algorithm is the use of the recently proposed meshless \ac{POD} \citep{tirelli2024meshless} to evaluate the similarity between snapshots in the KNN-PTV.
The features extracted from the meshless POD are less affected by random errors and thus offer a more accurate definition of similarity. \textcolor{black}{This combination results in a super-resolution methodology able to address all the limitations discussed in the previous paragraphs. The approach yields an analytical representation of the flow field while keeping computational costs at a reasonable level.}

The methodology is detailed in \S~\ref{sec.methodology} . The experimental dataset involves three-dimensional PTV measurements of a jet flow in air, described in \S~\ref{sec:jet}.  Lastly, the  assessment is discussed in \S~\ref{sec.results} in terms of statistics (\S~\ref{sub.stat}), spectra (\S~\ref{sub.spec}) and modal analysis (\S~\ref{sub.mod}). \textcolor{black}{Furthermore, \S~\ref{app} illustrates the benefits of the KNN-driven densification process, while \S~\ref{subsec.const} highlights the role of physical constraints in regularizing the regression.}

\section{Methodology}
\label{sec.methodology}

The workflow of the meshless KNN-PTV with c-RBFs is summarised in Fig.~\ref{Fig:flowchart}. As in the KNN-PTV proposed by \citet{tirelli2023end}, the cornerstone of the super-resolution strategy is the merging of particles belonging to different snapshots in order to artificially increase the particle density. Particles from different snapshots are merged only when a local similarity is identified. To this end, the domain is divided into subdomains and, for each of them and at each time instant, the algorithm searches for the most locally similar realizations within the data ensemble. If the flow fields in a given subdomain at different time instants are deemed sufficiently similar, their particles are merged to form a denser snapshot. The similarity is assessed using a local meshless \ac{POD} \citep{tirelli2024meshless}.

\textcolor{black}{The POD is employed exclusively as a low-dimensional statistical embedding to efficiently evaluate similarity between realizations and does not enter the reconstruction process. The instantaneous flow fields are reconstructed independently via constrained \acp{RBF} regression, without projection onto POD modes or any form of temporal reconstruction.}

The enriched particle distributions are then used to feed a constrained interpolator based on \acp{RBF} \citep{sperotto2022meshless}, yielding an analytical representation of the flow field. The algorithm consists of four main steps, outlined below and summarized in Algorithm~\ref{pseudo}. 

\vspace{0.2cm}

\textbf{Step 1: Particle detection} 

\vspace{0.2cm}
The first step involves extracting information from particles using conventional \ac{PTV} methods \citep{keane1995super} or more modern \ac{LPT} algorithms, such as \acused{STB} Shake-The-Box \citep[STB,][]{schroder20233d}. In the implementation presented in this manuscript, particle identification and pairing are performed with a traditional \ac{PTV} algorithm. However, employing a more accurate tracking algorithm, such as \ac{STB}, is anticipated to enhance the performance. For example, tracks obtained from \ac{STB} could be used to enforce temporal coherence in the regression process. However, minor improvements are expected for time-resolved sequences, in which well-assessed methods of pouring time resolution into space can be used instead \citep{sciacchitano2012multi,schneiders2014time,schneiders2016dense}.

\begin{algorithm*}[h!]
	\caption{Meshless KNN-PTV \vspace{0.1cm}}
	\begin{algorithmic}[1]
		\Require Data: $\{\bm{u}(\mathbf{X}^{(i)}, \mathbf{t}_i)\}$ for $i = 1, 2, \ldots, N_t$ (scattered in space and time).
		\Ensure High-resolution analytical flow fields.
		\vspace{0.1cm}
		\Statex\hrulefill
		\Statex \textit{\textbf{Step 1:} Particle detection}
		\Require particle image pairs or sequences.
		\Ensure Particle distribution. 
		\vspace{0.1cm}
		\For{each time step $\mathbf{t}_i$}
		\State Perform PTV \citep{keane1995super} or LPT \citep{schroder20233d}.
		\EndFor
		\Statex\hrulefill
		\Statex \textit{\textbf{Step 2:}  Local Meshless POD \citep{tirelli2024meshless}}
		\Require Data: $\{\bm{u}(\mathbf{X}^{(i)}, \mathbf{t}_i)\}$ for $i = 1, 2, \ldots, N_t$ (scattered in space and time).
		\Ensure Training set $\bm{\Theta} = \bm{\Psi}_r\bm{\Sigma}_r$. 
		\vspace{0.1cm}
		\For{each subdomain}
		\For {each time instant $t_k$}
		\State Compute analytical approximation through RBF: $\bm{\tilde{u}}(\bm{x}, \bm{t}_k) = \sum_{j=1}^{N_b^{(k)}} \bm{a}_j(\bm{t}_k) {\gamma}_j(\bm{x};\bm{X}_{c,j}^{(k)})$.
		\EndFor
		\State Compute temporal correlation matrix: $\mathbf{K} \in \mathbb{R}^{N_t \times N_t}$.
		\State Decompose  $\mathbf{K}$ through SVD: $\mathbf{K} = \mathbf{\Psi}\mathbf{\Sigma}^2\mathbf{\Psi}^T$
		\State Assemble training set $\bm{\Theta} = \bm{\Psi}_r\bm{\Sigma}_r$, with $r$ the number of modes that retain the $90\%$ of the energy.
		\EndFor
		\Statex\hrulefill
		\Statex \textit{\textbf{Step 3:}   Enriching snapshots and RBFs placing}
		\Require Data: $\{\bm{u}(\mathbf{X}^{(i)}, \mathbf{t}_i)\}$ for $i = 1, 2, \ldots, N_t$; local training set $\bm{\Theta}$.
		\Ensure Enriched snapshots.
		%\vspace{0.1cm}
		\For{each high-resolution snapshot needed}
		\For{each subdomain}
		\State Compute correlation matrix $\bm{\mathcal{S}} = \bm{K}_r \oslash (\bm{\kappa}\bm{\kappa}^T)$
			\State Compute the number of neighbours $k$ as the number of element in $\bm{\mathcal{S}}$ higher than a given threshold 
		\State Increase local particle density according to $k$ via KNN
		\EndFor
		\State Iterative agglomerative clustering to place collocation points and corresponding RBFs \citep{sperotto2022meshless}
		\EndFor
		
		\algstore{algo} % Store the algorithm here and continue on next page
	\end{algorithmic}
	\label{pseudo}
\end{algorithm*}

\begin{algorithm*}[t]
	\begin{algorithmic}  
		\algrestore{algo} % Restore numbering and continue
	
		\Statex \textit{\textbf{Step 4:}   RBF constrained weighted regression}
		\Require Enriched snapshots, collocation points, training set.
		\Ensure High-resolution analytical flow fields
		\State Compute weighting matrix $\mathbf{\Xi}$ as in Eq.~\eqref{eq.xi}
		\For{each high-resolution snapshot needed}
		\State Perform RBF weighted regression on the enriched particle distribution (solving the system in Eq.~\eqref{eq:linear_system})
		\EndFor
		\vspace{0.1cm}
	\end{algorithmic}
\end{algorithm*}

\vspace{0.2cm}

\textbf{Step 2: Local Meshless \ac{POD}} 

\vspace{0.2cm}
Meshless \ac{POD} is used in this step to assess the local similarity. The meshless \ac{POD} introduced by \citet{tirelli2024meshless} eliminates the dependence on an Eulerian grid to define this set of local feature dictionaries and mitigates the bias error introduced by the interpolation of scattered data. In the meshless \ac{POD}, the scattered velocity fields are approximated and replaced by analytical functions computed at each time instant $\bm{t}_k$. This approximation, denoted as $\bm{\tilde{u}}(\bm{x}, \bm{t}_k)$, is defined as a linear combination of \acp{RBF}:

\begin{equation}
	\bm{\tilde{u}}(\bm{x}, \bm{t}_k) = \sum_{j=1}^{N_b^{(k)}} \bm{a}_j(\bm{t}_k) {\gamma}_j\bigl(\bm{x};\bm{X}_{c,j}^{(k)}\bigr),
	\label{eq:func}
\end{equation}

\noindent where $\mathbf{a} \in \mathbb{R}^{N^{(k)}_b}$ is the vector collecting the weight that identifies the best approximation, and $\gamma_j(\bm{x};\bm{X}_{c,j}^{(k)})$ is the basis function j of the location positioned at the collocation point $\bm{X}_{c,j}^{(k)}$ available at the $k^{\text{th}}$ time instant. \textcolor{black}{Following \citet{tirelli2024meshless}, we use thin-plate spline (TPS) \citep{buhmann2000radial} as bases and an interpolative approach with the collocation points matching the sample points (that is, the number of bases $N^{(k)}_b$ coincides with the number of particles in the $k^{th}$ snapshot). This avoid hyper-parameters for this step, since the TPS have no shape factor and are defined as} 

\begin{equation} 
	\textcolor{black}{\gamma_j(\bm{x};\mathbf{X}_{c,j}^{(k)})=\gamma_j({r}(\bm{x};\mathbf{X}_{c,j}^{(k)})) = {r_j}^2\log({r_j}),}
	\label{eq.thinplate}
\end{equation}\textcolor{black}{with ${r}_j = \|\bm{x} - \bm{X}_{c,j}^{(k)}\|$ the distance from the  collocation (sample) point. }
\textcolor{black}{In addition, contrary to the regression approaches to velocity field reconstruction in Refs.~\cite{sperotto2022meshless,ratz2024meshless}, the interpolative formulation produces a positive definite linear system that can be solved directly, without tuning regularization parameters or forming normal equations. Rather than estimating an underlying trend by filtering out noise, interpolation enforces exact reproduction of the available samples while relying on the smoothness of the basis for physical plausibility. The rationale behind this choice is that the interpolated fields are only used to compute the temporal correlation matrix analytically, with noise mitigation left entirely to the inner product over the spatial domain of all velocity field approximations jointly across all variables. The temporal correlation matrix $\bm{K}$, defined in terms of continuous inner product \citep{lumley1967structure}, reads:}

\begin{equation}
	\label{Key}
	\mathbf{K}_{ij} = \frac{1}{|\Omega|} \int_\Omega \tilde{\bm{u}}^T(\bm{x}', \bm{t}_i) \tilde{\bm{u}}(\bm{x}', \bm{t}_j) \, d\bm{x}'\,,
\end{equation} with \textcolor{black}{$|\Omega|$ the area (in 2D) or volume (in 3D) of the spatial domain considered. The decomposition of the matrix $\bm{K}$ produces the temporal modes and their corresponding eigenvalues. 
	The computation of Eq.~\eqref{Key} is performed on the analytical approximation of the mean-shifted velocity fields of Eq.~\eqref{eq:func}, computed as detailed in Ref.~\citep{tirelli2023simple} .}

The analytic approximation provided by Eq.~\eqref{eq:func} allows for using quadrature methods to compute the integral in Eq.~\eqref{Key} with quadrature points, eliminating the need to interpolate data on a mesh. This reduces spatial modulation effects and enhances decomposition accuracy. The reader is referred to \citet{tirelli2024meshless} for more details on the meshless POD. The decomposition is applied to all the subdomains to extract local bases:

\begin{equation}
	\textcolor{black}{\bm{K} = \bm{\Psi}\bm{\Sigma}^2\bm{\Psi}^T }.  
\end{equation}

These are then used to construct a local feature training set, denoted as $\bm{\Theta} = \bm{\Psi}_r\bm{\Sigma}_r \in \mathbb{R}^{N_t \times r}$, consisting of the temporal modes and the eigenvalues of the subdomain truncated at rank $r$, which is the number of modes that retain the $90\%$ of the energy. 

\vspace{0.2cm}

\textbf{Step 3: Enriching snapshots and  \acp{RBF} placing} 
% \begin{figure}
	% \centering
	%     \begin{overpic}[scale=0.85,unit=1mm,grid]{step3.png}

		%   \put(40,45){\parbox{20mm}{\textcolor{black}{\textbf{NEIGHBOURS}}}}
		
		% \end{overpic}
	% \caption{}
	% \label{fig:step3}
	% \end{figure}

\begin{figure}[t]
	\centering
\begin{overpic}[width=0.95\textwidth, unit = 1mm]{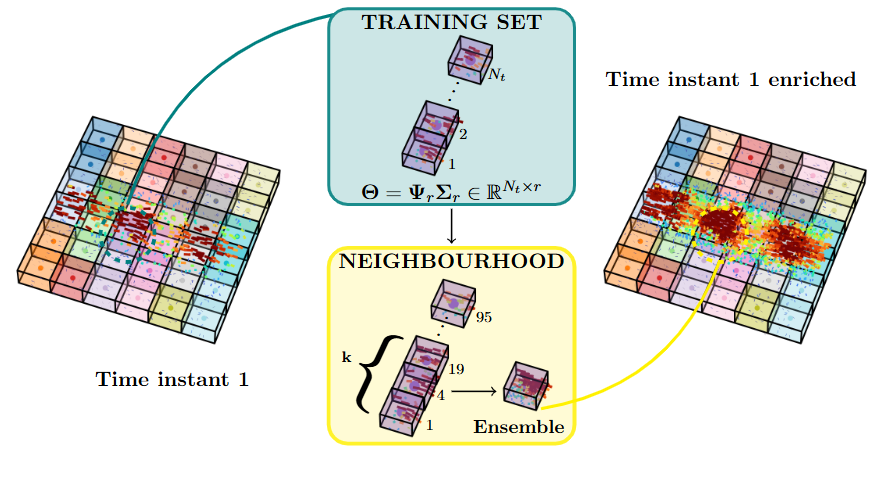}
\end{overpic}
	\caption{\textcolor{black}{Snapshot enrichment process: first build the training set $\bm{\Theta} = \bm{\Psi}_r\bm{\Sigma}_r \in \mathbb{R}^{N_t \times r}$ for each subdomain using meshless POD (cyan box); then select the $k$ nearest neighbors using the KNN algorithm (yellow box) and merge them. The process is repeated for all subdomains. The snapshot indexes indicated in the yellow box are arbitrary.}}
	
	\label{fig:enrichment}
\end{figure}

\vspace{0.2cm}
The third step involves determining the optimal number of neighbours for each subdomain based on the number of sufficiently (locally) similar snapshots in the dataset \textcolor{black}{and increasing the particle density according to this availability}.
The correlation between different time instants of each subdomain is employed as a metric. 

Using the continuous inner product, the correlation for the specific subdomain at two different time instants $i$ and $j$, is given by:

\begin{equation}
	\bm{\mathcal{S}}_{ij} = \frac{\int_\Omega \bm{u}(\bm{x},\bm{t}_i) \,\bm{u}(\bm{x},\bm{t}_j) \, d\bm{x}}{\sqrt{\int_\Omega \bm{u}(\bm{x},\bm{t}_i)^2 \,d\bm{x}}\,\sqrt{\int_\Omega \bm{u}(\bm{x},\bm{t}_j)^2 \,d\bm{x}}}.    
\end{equation}

This matrix can be obtained from the diagonal normalization of the matrix $\bm{K}$

\begin{equation}
	\bm{\mathcal{S}} = \bm{K} \oslash (\bm{\kappa}\bm{\kappa}^T),
\end{equation}

\noindent where $\bm{\kappa} \in \mathbb{R}^{N_t}$ collects the square root of the diagonal elements of $\bm{K}$ and $\oslash$ is the Hadamard division (entry by entry).

In this step, the similarity is assessed in the reduced-order version of $\bm{K}$, \textcolor{black}{referred here as $\bm{K_r}$}, obtained retaining only the $r$ modes accounting for the $90\%$ of the energy. The number of local neighbours $k$ is given by the number of elements in each row of $ \bm{\mathcal{S}}$ that exhibit a similarity higher than a certain threshold \textcolor{black}{(set for the remainder of the work to $0.75$)}.

This step replaces the need to create a reduced training set to search for the optimal number of neighbours $k$, which was the most computationally expensive part of the implementation in \citet{tirelli2023end}. In addition to reducing computational costs, the proposed approach also allows for the adaptation of $k$ for different time instants. It is worth noting that this step only determines the number of neighbours, not their positions. The positions are subsequently found using the KNN algorithm, which operates in the more refined feature space provided by the mesh-free modes. This process is repeated independently for each local subdomain, resulting in a comprehensive neighbour map in space and time. 

\textcolor{black}{The particle density is then artificially increased according to this map by merging particles from the $k$ nearest neighbours, i.e. the $k$ snapshots that, for a given time instant, exhibit a similarity above the predefined threshold within the specific subdomain. The procedure is illustrated in Fig.~\ref{fig:enrichment}. Once the training set $\bm{\Theta}$ and the optimal number of similar neighbours for each subdomain are defined, the KNN algorithm identifies the $k$ most similar snapshots in the training set. In the example proposed in Fig.~\ref{fig:enrichment}, the KNN explores the training set  (light blue box) to find the $k$ most similar snapshots to the analyzed one. The neighbourhood (yellow box) is built based on the Euclidean distance in the training set, and sorted accordingly. Particle density is then increased by ensemble merging from the $k$ nearest neighbours. This procedure is repeated for all subdomains and all snapshots to be reconstructed. }Subsequently, the collocation points needed for the regression are placed through iterative agglomerative clustering as in \citet{sperotto2022meshless}. 

\vspace{0.2cm}

\textbf{Step 4: RBF constrained weighted regression} 

\vspace{0.2cm}
The analytical high-resolution flow fields are obtained using the c-\acp{RBF} framework, as introduced by \citet{sperotto2022meshless}, with modifications to enhance compatibility with KNN-PTV. Unlike in Step $2$, this approach employs regressive \acp{RBF} with isotropic Gaussian basis functions $\varphi$:

\begin{equation} 
	\varphi_j(\bm{x};\mathbf{X}_{c,j}^{(k)}) = e^{-c^2 \|\bm{x} - \mathbf{X}_{c,j}^{(k)}\|^2_d}.
	\label{eq.gaussian}
\end{equation} 

The term ``isotropic'' refers to the fact that $c>0$ is the only shape parameter governing the basis function. 

In the framework of this work, the scattered data from \ac{PTV} or 
\ac{LPT} can be seen as samples of the analytical function of the $3$D flow field $\bm{u(x)}=(u(\bm{x}),v(\bm{x}),w(\bm{x}))$, which is to be approximated by the linear system: 

\begin{equation}
	\begin{aligned}
		\mathbf{u}(\mathbf{X}^{(k)}, t_k) &= 
		\begin{pmatrix}
			u(\bm{X}^{(k)}, t_k) \\
			v(\bm{X}^{(k)}, t_k) \\
			w(\bm{X}^{(k)}, t_k)
		\end{pmatrix}
		&\approx
		\begin{pmatrix}
			\bm{\Phi}_b(\bm{X}^{(k)}) & 0 & 0 \\
			0 & \bm{\Phi}_b(\bm{X}^{(k)}) & 0 \\
			0 & 0 & \bm{\Phi}_b(\bm{X}^{(k)})
		\end{pmatrix}
		\begin{pmatrix}
			\bm{a_u}(t_k) \\
			\bm{a_v}(t_k) \\
			\bm{a_w}(t_k)
		\end{pmatrix} \\[1em]
		&= \bm{\Phi}(\bm{X}^{(k)}) \bm{A}(t_k).
	\end{aligned}
	\label{eq:velocity_representation}
\end{equation}
where $\bm{\Phi}_b(\bm{X}^{(k)})\in \mathbb{R}^{{N_p(k)} \times {N_b(k)}}$ is the short-hand notation for $\bm{\Phi}_b(\bm{X}^{(k)} | \bm{X}^{(k)}_c, \bm{c}) $, obtained by evaluating the $N_b$ basis functions on the set of coordinates $\bm{X}^{(k)}$ with respect to the centers $\bm{X}^{(k)}_c\in\mathbb{R}^{N_p(k)\times 3}$ and the vector of shape factors $\bm{c}\in\mathbb{R}^{N_b}$. 

In the constrained formalism introduced by \citet{sperotto2022meshless}, both quadratic penalties and linear constraints can be incorporated into the regression process. \textcolor{black}{Penalties serve as soft constraints, promoting the minimization of specific quadratic terms without strictly enforcing them, thus maintaining the original dimensionality of the problem. In contrast, hard constraints are enforced explicitly through Lagrange multipliers $\boldsymbol{\lambda}$, which introduce additional unknowns into the system for each constraint at specified points. While hard constraints guarantee strict adherence to prescribed conditions, soft constraints provide flexibility by promoting, but not enforcing.}

\textcolor{black}{The use of linear penalties and quadratic constraints ensures that the associated augmented cost function, built to satisfy the Karush-Kuhn-Tucker optimality condition, remains in a quadratic form. The main interest in using both penalties and constraints is to strike a balance between computational cost and accuracy.} \textcolor{black}{A broader overview on the role of penalties and constraints and their impact on the regression is outlined in \S~\ref{subsec.const}.}

The novelty compared to the original implementation by \citet{sperotto2022meshless} lies in the application of regression to an enriched particle distribution. This requires introducing a weighting metric in the regression process to account for the fact that some particles do not originate from the current snapshot but are instead from their neighbours, which may have varying degrees of similarity to the current snapshot under consideration.

By slightly modifying the implementation in the open-source toolbox SPICY \citep{sperotto2024spicy,Mendez_BOOK}, the final cost function to be minimized, incorporating constraints such as Dirichlet and Neumann boundary conditions, as well as the solenoidal condition implemented both as penalty and constraint reads:

\begin{equation}
	\begin{split}
		\mathcal{J}(\bm{a}, \bm{\lambda}) &= \left\| \bm{U(X)} - \bm{\Phi({X})}\bm{A} \right\|^2_\Xi  +  \bm{\lambda}^T_D \left( \bm{D}(\bm{X}_D)\bm{A} - \bm{c}_D \right)   +  \bm{\lambda}^T_N \left( \bm{N}(\bm{X}_N)\bm{A} - \bm{c}_N \right)  \\
		&\quad  +  \bm{\lambda}^T_\nabla \left( \bm{D}_\nabla(\bm{X}_\nabla)\bm{A} \right)  + \alpha_\nabla \left\| \bm{D}_\nabla(\bm{X}_g) \bm{A} \right\|^2_2.
	\end{split}
	\label{eq: J_const_with_C}
\end{equation}

\noindent The first term $  \| \bm{U}(\bm{X}) - \bm{\Phi}(\bm{X}) \bm{A} \|_\Xi^2 $ is the weighted norm with respect to $\bm{\Xi}$. The diagonal weighting matrix $\bm{\Xi}$ is used here to penalise the information provided by other snapshots. The penalisation term is based on the distance in the local feature space $\bm{\Theta}$.
The matrix $\bm{\Xi}$ is computed as:

\begin{equation}
	\bm{\Xi} = e^{ -\left( \alpha\frac{\bm{D}}{\|\bm{\Theta}\|} \right)^2 } \in \mathbb{R}^{N_t\times N_t}.
	\label{eq.xi}
\end{equation} 

Similar to what is described in \citet{tirelli2023end}, this weighting coefficient accounts for the distances $\bm{D}$ in the feature space, normalized by the norm of the feature set and penalized by a factor $\alpha$.

In this step, physical constraints are also enforced to better suit the specific case study, as shown in Eq.~\eqref{eq: J_const_with_C}. The second, third and fourth terms on the right-hand side are the linear operators associated with the imposition of Dirichlet, Neumann and divergence-free conditions, applied to $\bm{X}_D \in \mathbb{R}^{N_D}$, $\bm{X}_N \in \mathbb{R}^{N_N}$ and $\bm{X}_\nabla \in \mathbb{R}^{N_\nabla}$ respectively, modelled as in \citet{sperotto2022meshless}.
The last term is a soft constraint used to penalize the violation of the divergence-free condition, acting on  $\bm{X}_g \in \mathbb{R}^{N_g}$ and weighted by the parameter $\alpha_\nabla \in\mathbb{R}^+$.

Another difference with the original framework, is that here only Gaussian basis functions have been employed. More complex \acp{RBF} can be easily integrated into the proposed framework. It is important to emphasize that the selection of basis functions does not affect the generality of the formulation.
Despite the well-known significant role that polynomial basis could play in terms of regularization of the regression and approximation of global behaviour, as shown in \citet{sperotto2022meshless}, it is also acknowledged that its effectiveness heavily relies on the expertise of the user. This is crucial as the correct scaling of the domain is required to position them optimally.
Inspired by the concept of the first KNN-PTV, aiming for an end-to-end tool that starts from raw images and yields output with minimal user intervention, the polynomial function has been excluded. This decision, while sacrificing a degree of accuracy, reduces the number of parameters to be selected. Future investigations will focus on identifying a set of bases that strikes the best compromise between accuracy and simplicity. 

The minimization of Eq.~\eqref{eq: J_const_with_C} with respect to the \acp{RBF} weights and the Lagrange multipliers associated with the constraints leads to the following system of equations:

\begin{equation}
	\centering
	\begin{pmatrix}
		\bm{\Gamma} & \bm{\Delta} \\
		\bm{\Delta}^T & \bm{0} 
	\end{pmatrix}
	\begin{pmatrix}
		\bm{A}  \\
		\bm{\lambda}
	\end{pmatrix}=
	\begin{pmatrix}
		\bm{b}_1 \\
		\bm{b}_2
	\end{pmatrix}\,,
	\label{eq:linear_system}
\end{equation} 

\noindent where $ \bm{\lambda} = \left(\bm{\lambda}_D, \bm{\lambda}_N, \bm{\lambda}_\nabla\right) \in \mathbb{R}^{n_\lambda} $, with $ n_\lambda = 3n_D + 3n_N + n_\nabla $ the total number of constraints. The matrices $ \bm{\Gamma} $ and $ \bm{\Delta} $, along with the vectors $ \bm{b}_1 $ and $ \bm{b}_2 $, are defined as follows:
\begin{subequations}
	\begin{equation}
		\bm{\Gamma} = 2 \mathbf{\Phi}^T(\mathbf{X}) \mathbf{\Xi}^T \mathbf{\Xi} \mathbf{\Phi}(\mathbf{X}) + 2 \alpha_\nabla \mathbf{D}_\nabla^T(\mathbf{X}) \mathbf{D}_\nabla(\mathbf{X}) \in \mathbb{R}^{3N_b \times 3N_b}  
	\end{equation}
	\begin{equation}
		\bm{\Delta} = \left(\bm{\Phi}^T_b\left(\bm{X}_D\right);\bm{N}^T_n\left(\bm{X}_N\right);\bm{D}^T_\nabla\left(\bm{X}_\nabla\right)\right) \in \mathbb{R}^{3N_b \times N_\lambda} 
	\end{equation}
	\begin{equation}
		\mathbf{b}_1 = 2 \mathbf{\Phi}^T(\mathbf{X}) \mathbf{\Xi}^T \mathbf{\Xi} \mathbf{U}(\mathbf{X}) \in \mathbb{R}^{3N_b}
	\end{equation}
	\begin{equation}
		\bm{b}_2  = \left(\bm{c}_D;\bm{c}_N;\bm{0}\right)
	\end{equation}
\end{subequations}

The final output of this step is a set of weights that enables the visualization of the analytical field on any grid while preserving the super-resolution achieved regardless of the underlying discretization.

\section{Experimental dataset: 3D jet flow} \label{sec:jet}

\begin{figure*}
	\centering
	\begin{overpic}[width=0.95\textwidth, unit = 1mm]{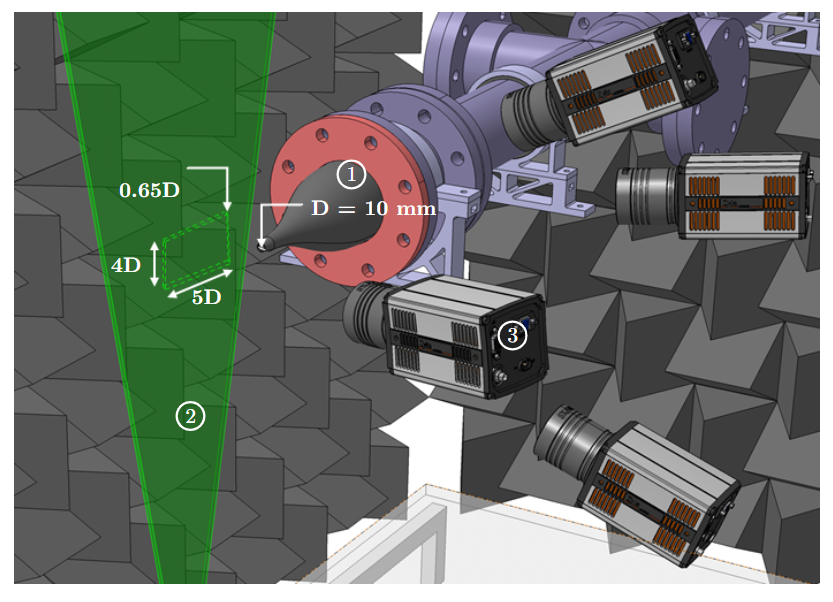}
		
	\end{overpic}
	\caption{Sketch of the experimental setup. (1) jet nozzle;  (2) Ng:Yag Quantel Evergreen laser; (3) ANDOR Zyla sCMOS $5.5$ MP camera.}
	\label{fig:setup}
\end{figure*}

The experimental validation aims to complete the process started with the first version of KNN-PTV. The assessment results reported in \citet{tirelliassessment} indicate that, although this algorithm produced encouraging outcomes, it struggled with larger interparticle spacing and increasing computational costs. This motivated the incorporation of \acp{RBF}, enabling the algorithm to adapt to $3$D scenarios while maintaining reasonable computational expenses.
For this reason, the experimental validation proposed in this work is carried out on the same $3$D jet flow of the above-mentioned paper.

The experiments are conducted in the jet flow facility located in the anechoic chamber of UC3M, as sketched in Fig~\ref{fig:setup}. The jet has a nozzle diameter of $10$ mm and is issued at a bulk velocity of $11.2$ m/s, resulting in a Reynolds number $\mathrm{Re} = 7,500$. DEHS particles, approximately $1 \, \mu\text{m}$ in diameter, are used to seed the jet. A Nd:Yag pulsed laser, with a maximum pulse power of $200$ mJ and a repetition rate of $15$ Hz, illuminates the particles. A domain of $50 \times 45 \times 6.5$ $mm^3$ (with the second dimension aligned along the axis of the jet) is imaged by four Andor Zyla sCMOS cameras ($5.5$ Mpx sensor, $6.5 \, \mu\text{m}$ pixel pitch). These cameras are equipped with objectives that have a focal length of $100$ mm and are set at $f_\#=11$. The four cameras are arranged in a cross-like configuration within the same plane, with an opening angle of $30^\circ$ in both directions.

The images undergo preprocessing using eigenbackground removal \citep{mendez2017pod} and a sliding minimum subtraction technique to set the background to zero. The self-calibration procedure introduced by \citet{wieneke2008volume} reduces the residual calibration error to below $0.1$ pixels. A tomographic reconstruction process \citep{elsinga2006tomographic} is performed using a multi-resolution method \citep{discetti2012fast} on a volume discretized with $28$ voxels/mm. The process involves three camera-simultaneous multiplicative algebraic reconstruction technique (cSMART) iterations on a $2\times$ binned configuration, followed by three additional cSMART iterations and three SMART iterations at the final resolution of $11$ voxels/mm. The cSMART is a modified version of the SMART procedure proposed by \citet{atkinson2009efficient} which uses the cameras sequentially.

Individual particles are identified in the reconstructed volume and paired, in accordance with the Tomo-\ac{PTV} principle established by \citet{novara2013particle}. A fast predictor is constructed using the sparse cross-correlation algorithm implemented by \citet{discetti2012fast}. A total of $1,000$ snapshots have been processed.

Approximately $10,000$ particles are accurately paired for each snapshot. This relatively low concentration is set to ensure a highly accurate reconstruction and a minimal occurrence of outliers, resulting in roughly $8$ vectors in a $64^3$ voxel volume. The complete distribution of vectors is used to create a reference ``ground truth'' field by weighting the moving average of the vector distribution with a Gaussian window, where the standard deviation is equal to $64/1.5$ voxels. The approaches evaluated here are tested on an artificially downsampled vector distribution, containing only $1,000$ particles distributed within the volume. Low-resolution fields are constructed using a moving average over windows of $128^3$ voxels, which contain, on average, $6.5$ particles in the sparse particle distributions.

The mesh-free flow fields are generated by distributing the RBF basis across eight levels of clustering, ensuring a minimum number of particles per Gaussian of $2$,$3$,$4$,$5$,$6$,$10$,$30$ and $50$ respectively, leading to an average value of $4,000$ basis for the enriched fields and almost the half for the traditional c-RBFs. Additionally, $1/10$ of the original particles are constrained to satisfy the divergence-free condition, which is further enforced as a penalty in the regression process.

The validation process is conducted within a reduced domain of interest, defined by the ranges $ 0.3 < x/D < 3.3 $, $ -1 < y/D < 1 $ and  $-0.2 < z/D < 0.2$. This selection ensures consistent particle coverage and well-converged results throughout the analyses.

\section{Results} \label{sec.results}
This section presents the validation of the algorithm on the experimental $3$D jet flow. The results are compared with the following.

\begin{itemize}
	\item \textbf{PIV IW = $128$}: represents the standard approach in the field, obtained via a moving average with an interrogation window (IW) size of $128$ voxels;
	\item \textbf{KNN-PTV}: the first version of the algorithm as proposed by \citet{tirelli2023end}, included to highlight the improvements introduced by the meshless paradigm;
	\item \textbf{c-\ac{RBF}}: meshless regression as in \citet{sperotto2022meshless}; it isolates the benefits of introducing particles from other snapshots.
\end{itemize}

The comparison of the meshless KNN-PTV against these approaches aims to demonstrate the advantages of the proposed combination. On one hand, the ensemble approach introduced by KNN-PTV enables higher spatial resolution, which is further enhanced and preserved through the analytical approximation of \acp{RBF}. On the other hand, the fully meshless nature of the method relies solely on particle positions, effectively avoiding modulation effects introduced by discretization on Eulerian grids at any step. \textcolor{black}{Notably, among the methods presented in the proposed comparison, the meshless KNN and c-RBfs are the only ones that incorporate physical constraints, which act as a form of regularization by enforcing physically consistent behavior, particularly helpful in regions with low particle density.} %where purely data-driven methods are more prone to overfitting or physically inconsistent results, as will be discussed in \S~\ref{subsec.const}.}

The quantitative assessment is carried out across three key aspects: statistical, spectral, and modal analyses.
The statistical analysis in \S\ref{sub.stat} includes ensemble and instantaneous statistics. For ensemble statistics, an additional reference is introduced: the EPTV approach developed by \citet{aguera2016ensemble}, using a bin size of $64$ voxels. This method represents the state-of-the-art algorithm for ensemble statistics in \ac{PIV} and serves as a further benchmark for comparison. For the spectral analysis, presented in \S\ref{sub.spec}, the goal is to evaluate and compare the frequency modulation introduced by all methods.
Finally, in terms of modal analysis, presented in \S\ref{sub.mod}, the goal was to evaluate the consistency of spatial modes as well as the convergence performances at varying ranks. 
\textcolor{black}{A qualitative analysis is proposed in \S~\ref{app}, highlighting the benefits of a densified particle distribution for evaluating derivatives of the velocity field (and, consequently, related quantities such as pressure), while also discussing the sources of the observed discrepancies.}
\textcolor{black}{Lastly, an ablation study on the role of constraints and penalties in the proposed methodology is presented in \S~\ref{subsec.const}. This subsection also provides the theoretical background needed to understand their formulation and practical implementation.}

In the analysis that follows, the velocity fields of the PIVs and KNN-PTV are filtered using the criterion proposed by \citet{raiola2015piv}. In contrast, the RBF-based approaches did not benefit significantly from this filtering, likely because the constrained regression already regularizes the fields. Therefore, in these cases, the filter has not been applied.

\subsection{Statistical analysis}\label{sub.stat}
\begin{figure*}[t]
\centering
\begin{overpic}[width=0.95\textwidth, unit = 1mm]{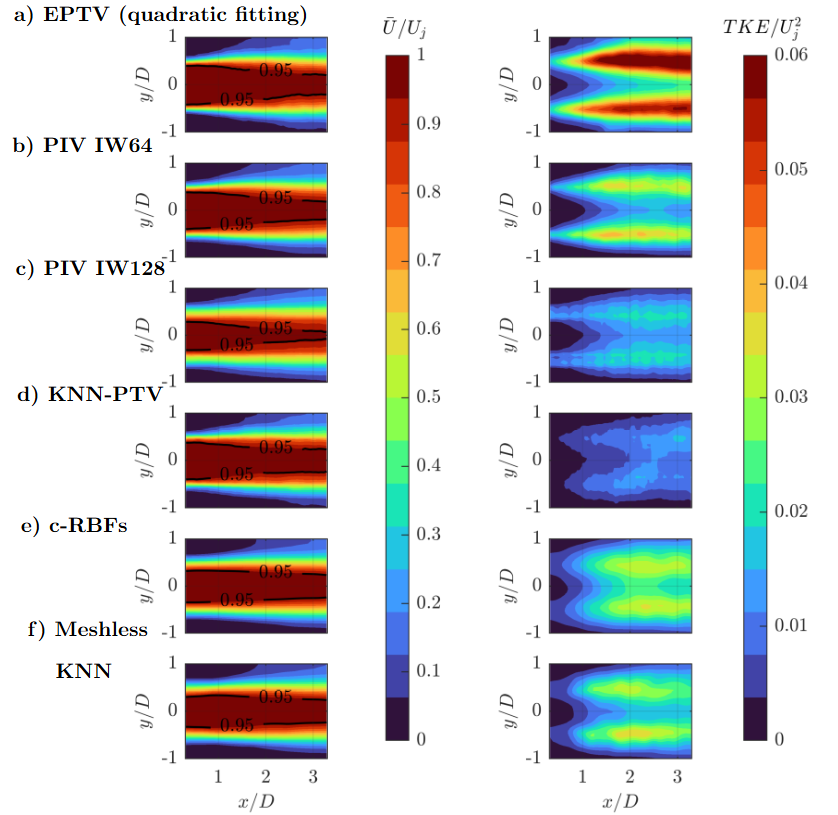}
	
\end{overpic}
\caption{Mean velocity field (left column) and resultant standard deviation $\mathbf{\sigma}$ squared (right column): a) Ensemble averaging \citep{aguera2016ensemble} with bin size $64$ voxels, b) \ac{PIV} with interrogation window of $64$ voxels, c) \ac{PIV} with interrogation window of $128$ voxels, d) KNN-PTV, e) c-RBFs and f) meshless KNN-PTV. Reference plane: $z/D = 0$. In black isolines for \textcolor{black}{$\bar{U}/U_j = 0.95$.} }
\label{fig. mean_jet}
\end{figure*}

\begin{figure*}[t]
\centering
\begin{overpic}[width=0.95\textwidth, unit = 1mm]{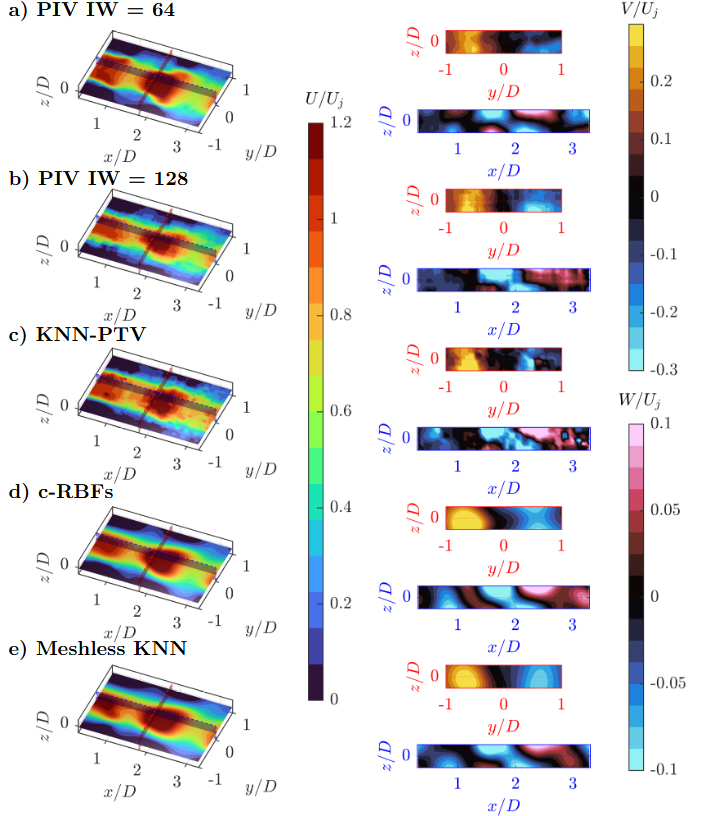}

\end{overpic}
\caption{Instantaneous streamwise (first column), spanwise (second column, top) and crosswise (second column, bottom) velocity field contours for the middle planes: a) reference \ac{PIV} with interrogation window of $64$ voxels, b) \ac{PIV} with interrogation window of $128$ voxels, c) KNN-PTV, d) c- RBFs and e) meshless KNN-PTV.   }
\label{fig.jet}
\end{figure*}

First- and second-order statistics are reported in Figure \ref{fig. mean_jet}. The comparison panel depicts the mean flow along the streamwise direction on the left and the turbulent kinetic energy (TKE) on the right, both in the plane $z/D = 0$ and normalized with the bulk velocity $U_j$. 

In addition to the reference PIV with IW = $64$ voxels, only for the comparison of ensemble statistics, the EPTV approach developed by \citet{aguera2016ensemble}, with a bin size of $64$ voxels (Fig.~\ref{fig. mean_jet}.a-left) is here used as further reference. The main differences in the mean flow arise from the analysis of the core region, highlighted by the black isolines at $\bar{U}/U_j = 0.95$: the PIV with IW = $128$ voxels (Fig.~\ref{fig. mean_jet}.b-left) is the only one whose core extension is shorter. This is probably due to the low availability of particles combined with the large moving average window that is over-filtering the field. The RBF-based approaches (Fig.~\ref{fig. mean_jet}.e, f-left) exhibit a contraction of the region slightly less pronounced than the references (Fig.~\ref{fig. mean_jet}.a,b-left) and KNN-PTV (Fig.~\ref{fig. mean_jet}.d-left).

The main discrepancies emerge from the comparison of the TKE plots. The EPTV exhibits the highest peaks in the shear layer (Fig.\ref{fig. mean_jet}.a-right). However, in the PIV with IW = $64$ voxels (Fig.\ref{fig. mean_jet}.b-right), despite being computed with the same number of particles ($10,000$), these peaks are smoothed out due to the larger window used for the moving average. Similarly, the PIV with IW = $128$  voxels (Fig.~\ref{fig. mean_jet}.c-right) shows a comparable pattern but with even more filtering, resulting from the combination of lower particle availability and a larger interrogation window. The KNN-PTV (Fig.~\ref{fig. mean_jet}.d-right), while producing an accurate mean field, has the poorest performance in terms of TKE. This is attributed to the inability of the algorithm to capture the smallest fluctuations. Three primary factors contribute to this: low particle availability in each snapshot, limited number of samples, and large interparticle spacing. Together, these factors limit the ability of KNN-PTV to find close neighbours for merging, thereby failing to resolve the smallest scales. On the other hand, the mean flow remains unaffected because the largest scales, which dominate the mean flow, are successfully captured. Lastly, the RBF-based methods (Fig.~\ref{fig. mean_jet}.e, f-right) recover the majority of the energy. Notably, the addition of particles provided by the KNN offers a slight boost to the already well-converged results of the c-RBFs. This enhancement enables the placement of smaller but well-supported Gaussian bases, thereby facilitating the accurate modelling of the smallest scales. \textcolor{black}{These findings are confirmed by the cosine similarities $\cos\theta$ reported in Tab.~\ref{tab}, computed with respect to the EPTV statistics.}

\begin{table}[h!]
\centering
\begin{tabular}{lcc}
	\hline
	\textbf{Method} & \textbf{$\cos\theta$ (Mean)} & \textbf{$\cos\theta$ (TKE)} \\
	\hline
	PIV64 & 0.999 & 0.985 \\
	PIV128 & 0.996 & 0.967 \\
	KNN-PTV & 0.999 & 0.958 \\
	c-RBFs & 0.998 & 0.967\\
	Meshless KNN & 0.998 & 0.976 \\
	\hline
\end{tabular}
\caption{\textcolor{black}{Cosine similarity computed with respect to the EPTV.} }
\label{tab}
\end{table}

A qualitative comparison of the instantaneous field is shown in Fig.~\ref{fig.jet}. In all test cases, contours of the instantaneous streamwise (left column), spanwise (right column - top) and crosswise (right column - bottom) velocity field for the corresponding middle planes ($z/D = 0$, $y/D = 0$ in blue and $x/D = 1.7$ in red)  are displayed. %In yellow, positive values of $Q$-criterion visualization \citep{hunt1988eddies} are reported, slightly smoothed using the Savitzky-Golay filter for visualization purposes \citep{savitzky1964smoothing}.

Comparing the \ac{PIV} results with an interrogation window of $128$ voxels (Fig.~\ref{fig.jet}.b) to the reference (Fig.~\ref{fig.jet}.a), the former appears as a lower resolution version due to the lack of particles and larger moving averaging windows, that implies high smoothing effects on the field.
The KNN-PTV implementation (Fig.~\ref{fig.jet}.c) seems to recover smaller scales more effectively at first glance but suffers more noise contamination.
The introduction of \ac{RBF} seems to help regularize the flow field, making it appear smoother, as evident in Fig.~\ref{fig.jet}.d. The \textcolor{black}{spatial resolution} is increased thanks to the availability of particles borrowed from other snapshots, as shown in Fig.~\ref{fig.jet}.e.

\begin{figure}[t]
\centering
\begin{overpic}[width=0.95\textwidth, unit = 1mm]{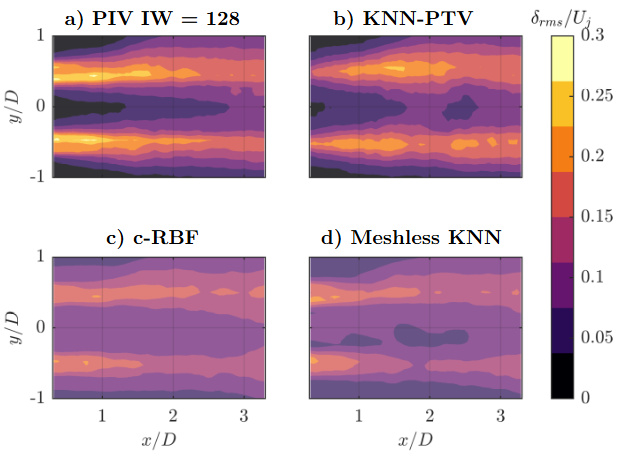}

\end{overpic}
\caption{Spatial distribution of $\delta_{RMS}$ normalized with $U_\infty = 12$ m/s for the plane at $z/D = 0$:  a) \ac{PIV} with interrogation window of $128$ voxels, b) KNN-PTV, c) c-RBFs and d) meshless KNN. }
\label{fig.jet_rms}
\end{figure}

These qualitative findings are confirmed by the error maps shown in Fig.~\ref{fig.jet_rms}. Here the root mean square error $\delta_{RMS}$ has been employed as a metric, normalized with $U_j$ and computed as:

\begin{equation}
\delta_{RMS} = \frac{||\mathbf{u}-\mathbf{u}_{ref}||}{\sqrt{N_t}\,U_j},
\label{eq:rms}
\end{equation}

\noindent where the reference is always the \ac{PIV} with IW = $64$ voxels. 

These maps, evaluated at $z/D = 0$, reveal that the highest errors occur near the shear layer region, reflecting the patterns observed in the TKE plots of Fig.~\ref{fig. mean_jet}-right. In Fig.~\ref{fig.jet_rms}.a, the error peaks are concentrated in the shear layer closer to the exit, between $x/D = 0.3$ and $1$, while the core region remains less affected, likely due to the lower variability in velocity fluctuations. These peaks are smoothed in the KNN-PTV results (Fig.~\ref{fig.jet_rms}.b), thanks to the artificial increase of particle density, which aids in capturing smaller fluctuations. The introduction of RBFs further decreases the peaks and reduces the average error, although this comes with slightly elevated errors in the core region (Fig.~\ref{fig.jet_rms}.c). The combination of these two methodologies achieves the best balance and overall performance, as confirmed by the spatial average of these maps $\langle\delta_{RMS}\rangle$ in Tab.~\ref{table:error}. \textcolor{black}{Here, the computed variance of the squared error $\mathrm{Var}(\delta^2)$, together with the maximum absolute error values 
$\text{max}(\delta_{RMS})$, provides a quantitative confirmation of the spatial error distributions observed in the RMS error maps. Specifically, the meshless KNN method demonstrates a notably more compact and homogeneous error distribution across the domain. This reduced variance indicates that the errors are more consistently distributed, with fewer extreme local deviations or sharp error peaks, especially when compared to the non-meshless approaches.}

\begin{table}[t]
\centering
\textcolor{black}{
	\begin{tabular}{||c|c|c|c|c||}
		\hline
		& PIV (IW = 128 px) & KNN-PTV & c-RBF & Meshless KNN-PTV \\ 
		\hline
		$\langle\delta_{\text{RMS}}\rangle$ & 0.1267 & 0.1237 & 0.1140 & 0.1030 \\
		\hline
		$\mathrm{Var}(\delta^2)$ & 0.1163 &  0.0693 &   0.0394 & 0.0316 \\
		\hline
		$\text{max}(\delta_{RMS})$ &  1.4109 &  1.0427 &    0.9179 &  0.8216 \\
		\hline
	\end{tabular}
}
\caption{Spatial average of the root mean square error $\langle\delta_{\text{RMS}}\rangle$, variance of the squared error $\mathrm{Var}(\delta^2)$ and maximum absolute error $\text{max}(\delta_{RMS})$ computed for: PIV with interrogation window of $128$ voxels, KNN-PTV, c-RBF and meshless KNN-PTV, all normalised with the bulk velocity $U_j$ (and its square for the variance). }
\label{table:error}
\end{table}

\subsection{Spectral analysis}\label{sub.spec}
\begin{figure}[t]
\centering
\begin{overpic}[scale=1,unit=1mm]{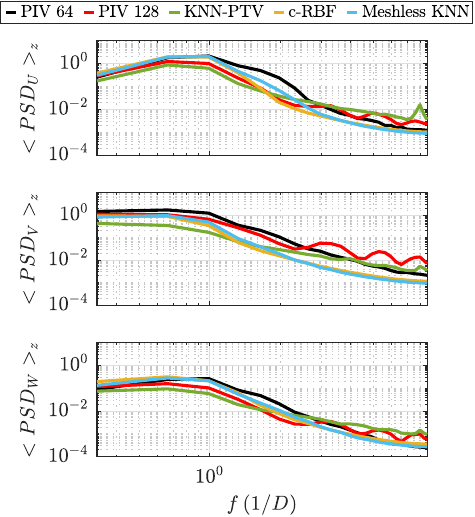}
	
\end{overpic}
\caption{Power spectral density comparison at $y/D = 0.5$, averaged at different $z/D$: streamwise component $U$ (first row), spanwise component $V$ (second row), and crosswise component $W$ (third row). The methodologies compared include: PIV IW = $64$ (reference, black); PIV IW = $128$ (red); KNN-PTV (light green); c-RBF (yellow); and meshless KNN (light blue). }
\label{fig.jet_spectra}
\end{figure}

The streamwise velocity spectra are presented in Fig.~\ref{fig.jet_spectra} in terms of Power Spectral Density (PSD). The PSD is computed for a velocity profile at $y/D = 0.5$ within the shear layer, evaluated at $11$ equispaced stations along the $z$-direction and then averaged. This analysis is performed independently for all three velocity components.

The energy spectra of the reconstructed fields are compared with the reference PIV (in black). In general, the meshless KNN-PTV (light blue line) provides results closest to the reference. The c-RBFs (yellow line) exhibit a similar pattern, although performing slightly worse than the meshless KNN-PTV in the range $1/D < f < 3/D$, particularly evident in the streamwise component analysis. This is due to the filtering effect of using a larger kernel for the RBFs, while the meshless KNN-PTV can use smaller kernels due to the artificially-increased particle image density. The KNN-PTV as in the implementation by \citet{tirelli2023end},  depicted with a light green line, follows the reference reasonably well up to a certain frequency ($\approx 3/D$), displaying more stable behaviour compared to the PIV with an interrogation window of $128$ voxels (red line), that tends to an oscillatory behaviour. However, the spectra deviate at high frequencies due to increased noise. \textcolor{black}{This early decay suggests a reduced ability to capture the smaller-scale structures of the flow.} 

\textcolor{black}{In contrast,} the proposed meshless blending of KNN-PTV with constrained RBF regression exhibits enhanced robustness at high frequencies, primarily due to the regularization introduced by the constrained regression and the avoidance of modulation effects caused by moving averaging. Furthermore, the RBF-based regression contributes to a greater resistance to high-frequency noise contamination when compared to other methodologies. \textcolor{black}{As a result, the method preserves the expected spectral decay over a wider frequency range, indicating improved spatial fidelity and a better ability to resolve fine-scale flow structures.}

\subsection{Modal analysis}\label{sub.mod}

\begin{figure*}[t]
\centering
\begin{overpic}[width=0.95\textwidth, unit = 1mm]{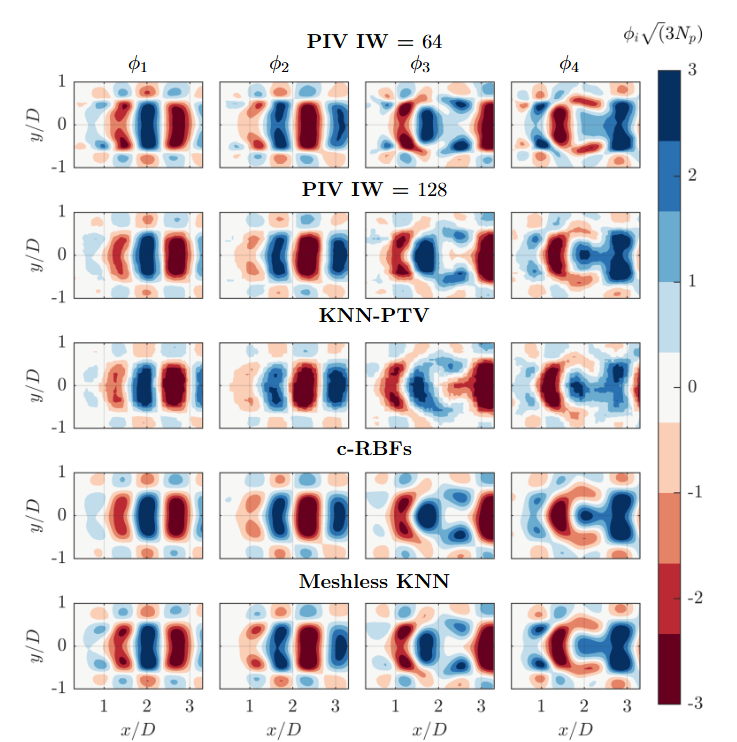}
	
\end{overpic}
\caption{First $4$ spatial modes $\phi$ normalized with their standard deviation. First row: reference PIV with IW = $64$; second row: PIV with IW = $128$; third row: KNN-PTV; fourth row: c-RBFs; fifth row: Meshless KNN. }
\label{fig.jet_phi}
\end{figure*}

\begin{figure}[t]
\centering
\begin{overpic}[scale=1,unit=1mm]{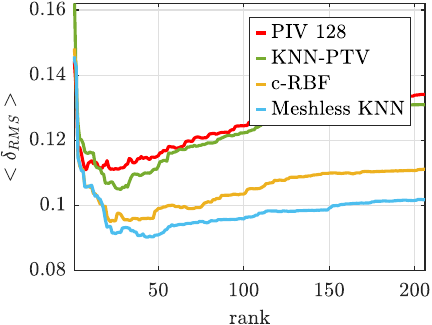}
\end{overpic}
\caption{Root mean square  error $<\delta_{RMS}>$ for varying rank $r$. The methodologies compared include: PIV with IW = $128$ (red), KNN-PTV (green), c-RBFs (yellow) and meshless KNN (light blue). }
\label{fig.jet_rms2}
\end{figure}

The last part of the assessment involves a modal analysis conducted using POD via the method of snapshots \citep{sirovich1987turbulence}. The reference data, along with those reconstructed through the different methodologies, have been decomposed to extract the POD modes.

Figure~\ref{fig.jet_phi} displays the streamwise component of the first four dominant spatial modes $\phi_i$, depicted in the plane $z/D = 0$ and normalized by their standard deviation ($\sqrt{3N_p}$). 
The first two modes are associated with the convective motion of the vortex rings forming in the free shear layer \citep{violato2013three}. These vortex rings are a characteristic feature of jet flows, particularly in transitional regimes, sustained by the Kelvin-Helmholtz mechanism of shear-layer instability. As these structures travel downstream, they tend to interact and enter a precessing motion in pairs, often referred to as ``leapfrogging'' \citep{schram2003aeroacoustics}.  
This leapfrogging mechanism is represented in the third and fourth modes, $\phi_3$ and $\phi_4$. During this process, the azimuthal modes grow rapidly, leading to the distortion of vortex filaments and their eventual breakup into smaller, three-dimensional fluctuations.

From a qualitative perspective, all the benchmark methods exhibit good agreement with the reference modes, accurately capturing the aforementioned mechanisms. As a general trend, the RBF-based approaches stand out as the methods that replicate these patterns with the highest accuracy, particularly in capturing the smallest scales, while the KNN-PTV appears to be the most affected by noise contamination. In modes $1$ and $2$, PIV with IW = $128$ and KNN-PTV fail to accurately model the first vortex pairs at $x/D = 1$. On the other hand, the RBF-based methods achieve the closest reconstruction, with the meshless KNN being also able to accurately reconstruct the shape of the modes and their velocity peaks. All of them are not able to capture the small vortices at the beginning of the region of interest. This is primarily due to the lack of particles and the dimensions of such structures, which make their recovery very challenging with traditional POD. The modulation effect on the data, once stored on an Eulerian grid, further hinders this process. A potential improvement in this regard could be achieved using meshless POD \citep{tirelli2024meshless}, but this lies beyond the scope of the present work. Modes $3$ and $4$ confirm the findings of the previous ones: the meshless KNN exhibits the closest modes to the reference ones, successfully recovering the smallest scales and velocity peaks while maintaining the same shape as the reference modes.

The qualitative findings are confirmed by the reconstructed flow fields for varying rank $r$, reported in Fig.~\ref{fig.jet_rms2}. The reconstructions are compared in terms of RMS error (computed as in Eq.~\eqref{eq:rms}) at different ranks, where the meshless KNN achieves the highest accuracy across all ranks.

\subsection{\textcolor{black}{Velocity derivatives and pressure estimation via KNN-driven densification}} \label{app}

\begin{figure*}[t]
\centering
\begin{overpic}[width=0.95\textwidth, unit = 1mm]{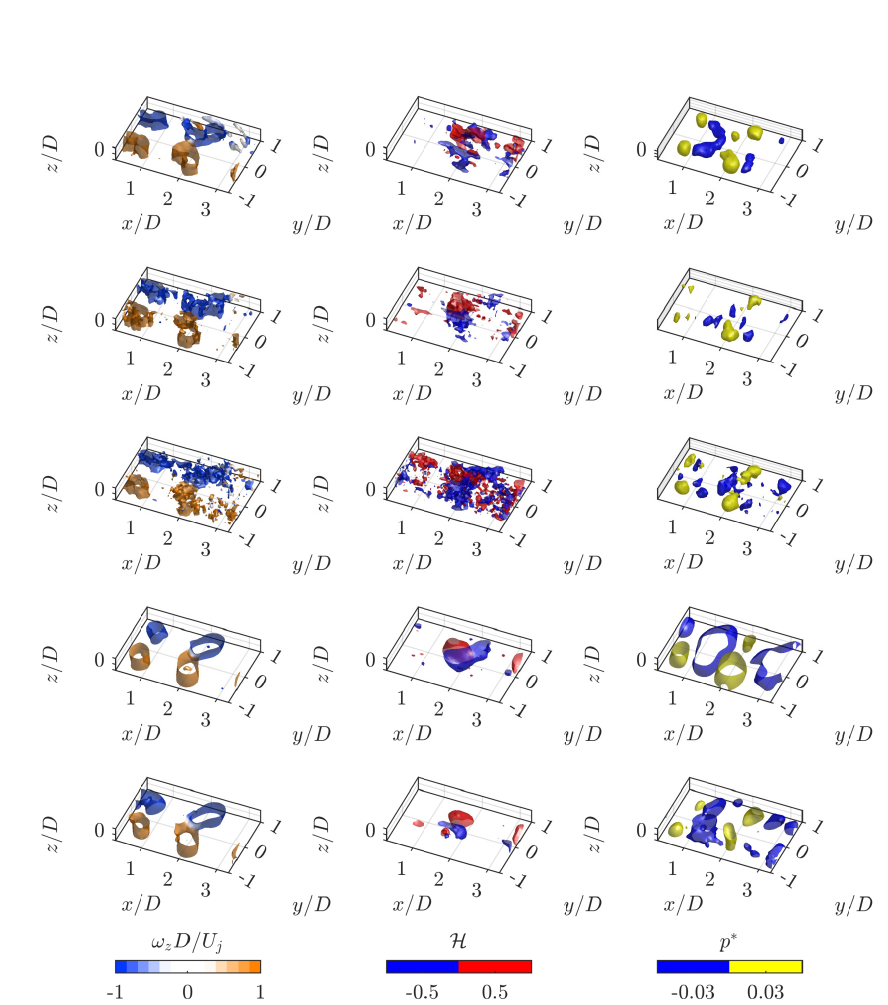}
	\put(0,130){\parbox{50mm}{  \flushleft \textnormal{\textbf{ a) PIV IW = 64}}}}
	\put(0,103.5){\parbox{50mm}{  \flushleft\textnormal{\textbf{ b) PIV IW = 128}}}}
	\put(0,80){\parbox{50mm}{ \flushleft \textnormal{\textbf{ c) KNN-PTV}}}}
	\put(0,56){\parbox{50mm}{ \flushleft \textnormal{\textbf{ d) c-RBFs }}}}
	\put(0,33){\parbox{50mm}{  \flushleft\textnormal{\textbf{ e) Meshless KNN }}}}
\end{overpic}
\caption{\textcolor{black}{Q-criterion visualization for positive values of $Q$, coloured by the crosswise vorticity $\omega_zD/U_j$  (first column); isocontours of positive and negative helicity $\mathcal{H}$  (second column); isocontours of positive and negative normalized pressure $p^*$ (third column). Depicted cases: a) reference \ac{PIV} with interrogation window of $64$ voxels, b) \ac{PIV} with interrogation window of $128$ voxels, c) KNN-PTV, d) c- RBFs and e) meshless KNN-PTV.}}
\label{fig.Q}
\end{figure*}

\textcolor{black}{
Once the analytical approximation of the flow field is available, its spatial derivatives and derived quantities can be obtained by simply computing the derivatives of the basis functions (see Appendix II).
The novelty of the proposed approach lies in the possibility of computing such derivatives on a scattered enriched field, obtained by leveraging the information provided by selected neighbors. %This artificial and targeted increase of particle density preserves the benefits of a denser distribution, without incurring the drawbacks discussed at the beginning. 
}

\textcolor{black}{
Figure~\ref{fig.Q} reports three quantities that involve the derivatives of the velocity field: the second invariant ($Q$) of the velocity gradient tensor, the helicity $\mathcal{H}$ and the pressure field. The snapshot represented here is the same as Fig.~\ref{fig.jet}, as well as the methodologies compared. The first column reports positive values of $Q$-criterion visualization \citep{hunt1988eddies}, coloured by the crosswise vorticity ${\omega_z}$, normalized with $D/U_j$. The $Q$-criterion is a widely used method for the identification of vortical structure in the flow field, computed as:
}

\begin{equation}
\mathbf{Q} = \frac{1}{2} \left( || \mathbf{\Omega}||^2_F - ||\mathbf{S}||^2_F \right),
\end{equation}

\noindent \textcolor{black}{
with $\mathbf{\Omega}$ representing the vorticity tensor (antisymmetric part of the velocity gradient tensor $\mathbf{\nabla U}$) and $\mathbf{S}$ the rate of strain tensor (symmetric part of $\mathbf{\nabla U}$). Positive values of $\mathbf{Q}$ represent a predominance of the vorticity over the strains, i.e vortical structures. 
}

\textcolor{black}{
The distribution of $\omega_z$ highlights the presence of counter-rotating vortices at different stages of their natural development. The PIV field with $IW = 128$ shows good agreement with the reference, although it appears noisier and loses the smallest details. As explained before, the larger windows and the lower seeding emphasize it. This noise amplification in the derivatives is particularly evident for the case of the KNN-PTV reported in Fig.~\ref{fig.Q}.c, which consistently exhibited a higher noise contamination throughout the analyses previously carried out. Nevertheless, the use of a common grid and the same finite-difference scheme for vorticity estimation of the reference helps to mitigate these discrepancies. 
}

\textcolor{black}{
On the other hand, the meshless methods exhibit smoother fields, while losing some details; %As reported in \citet{raffel2018particle}, this smoothing effect can be probably ascribed to the fact that these methods are based on a least-square regression. 
among the two, the meshless KNN seems to be the one that closely follows the reference.
}

\textcolor{black}{
Interestingly enough, these two are the only ones recovering the small region of near-zero vorticity emerging between adjacent vortices, marking shear interfaces where rotational motions cancel out and potentially indicating zones of vortex interaction or pairing. These findings can also be interpreted in terms of helicity \citep{moreau1961constantes}, computed as:
}

\begin{equation}
\mathcal{H} = \mathbf{u}~\cdot~\bm{\omega} .
\end{equation}

\textcolor{black}{
This scalar quantity provides insight into the alignment between the velocity and vorticity fields: positive values indicate alignment (i.e. the vectors point in the same direction), corresponding to a right-handed helical structure, whereas negative values indicate anti-alignment, associated with a left-handed helix. In the second column of Fig.~\ref{fig.Q}, $ \mathcal{H} $ is computed using dimensionless quantities, with positive and negative isosurfaces highlighted in red and blue, respectively.
}

\textcolor{black}{
All the plots show a localized core at the previously discussed shear interfaces. This observation reinforces earlier findings, as the interaction between counter-rotating vortices in this region induces flow stretching, resulting in localized helicity. Similar considerations apply as in the previous plots: noisier fields tend to be amplified in the derived quantities, complicating interpretation. In contrast, the meshless methods, although subject to intrinsic smoothing, yield a more coherent representation of the helical structures. The meshless KNN seems to be the most consistent among the two.
}

\textcolor{black}{
The last column of Fig.\ref{fig.Q} presents a qualitative comparison of the positive and negative isosurfaces of estimated normalized pressure fields:
}

\begin{equation}
p^* = \frac{p}{\rho U_j^2},
\end{equation}

\noindent \textcolor{black}{
where $p$ is the estimated pressure and $\rho$ the density of the air.
}

\textcolor{black}{
For the gridded approaches, the pressure is obtained by integrating the Poisson equation using the Successive Over-Relaxation method \citep{saad2003iterative}, applying Dirichlet boundary conditions at the spanwise domain boundaries. The meshless methods, on the other hands, employ constrained regression, as in \citet{sperotto2022meshless},  with the same boundary conditions. As previously discussed for the velocity fields, the meshless KNN method applies the same weighting scheme to account for the contribution of particles from different neighbors.
}
\textcolor{black}{
The PIV with $IW = 128$ appears to lose many of the details of the reference, whereas the KNN-PTV method is able to reconstruct the pressure field with good fidelity. This accuracy is likely supported by the fact that it is computed using the same numerical scheme as the reference, helping to mitigate discrepancies. In contrast, the traditional constrained RBF approach tends to oversmooth the flow field, leading to a less accurate reconstruction. Although it employs the same configuration used for the meshless KNN, the latter provides the best overall performance.
}

\textcolor{black}{
It is possible that the c-RBF approach would benefit from a dedicated configuration and more extensive hyperparameter tuning, an effort that was not necessary for the velocity field reconstruction. In the case of pressure, the involvement of second-order derivatives makes the reconstruction more sensitive to the setup of the RBFs, likely accentuating differences between methods. Interestingly, the optimal configuration for pressure was found to differ from that used for the velocity field, reinforcing this observation. Furthermore, the penalty terms used to enforce divergence constraints in the velocity field needed to be reduced to prevent oversmoothing when preparing the pressure regression.
}

\subsection{\textcolor{black}{Effects of physical constraints and penalties}}
\label{subsec.const}
%\textcolor{red}{This subsection focuses on the role and the impact that physical constraints and penalties play in the regression.}

\begin{figure}[t]
\centering
\begin{overpic}[width=0.95\textwidth, unit = 1mm]{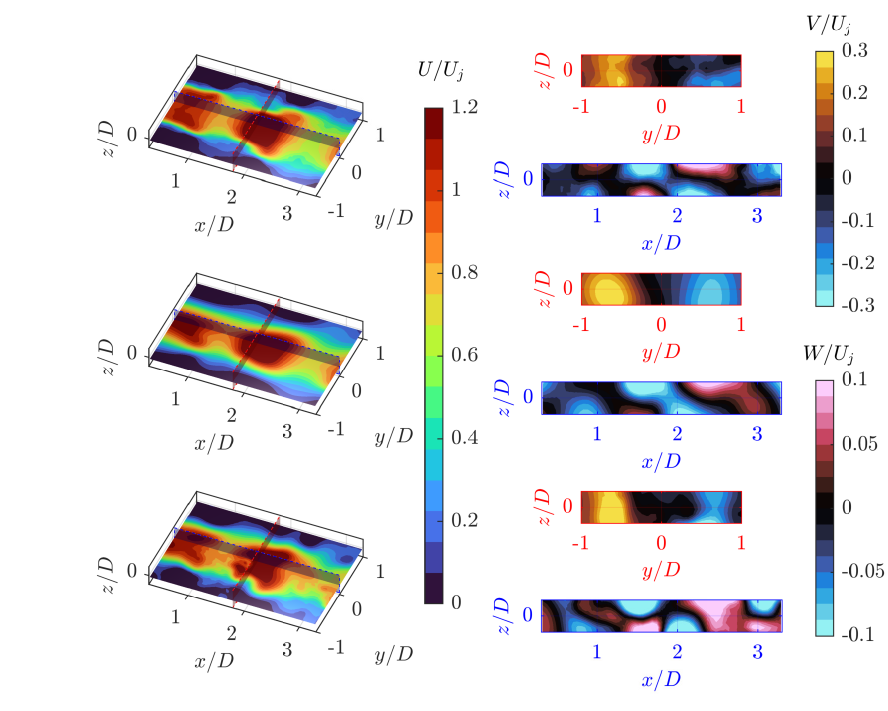}
	\put(0,93){\parbox{50mm}{  \flushleft \textnormal{\textbf{ a) PIV IW = 64}}}}
	\put(0,62){\parbox{50mm}{  \flushleft \textnormal{\textbf{ b) Constrained}}}}
	\put(0,32){\parbox{50mm}{  \flushleft \textnormal{\textbf{ c) Unconstrained}}}}
	
\end{overpic}
\caption{Instantaneous streamwise (first column), spanwise (second column, top) and crosswise (second column, bottom) velocity field contours for the middle planes: a) reference \ac{PIV} with interrogation window of $64$ voxels, b) meshless KNN-PTV with penalties and constraints and c) meshless KNN-PTV without penalties and constraints.   }
\label{fig.ConstvsUnc}
\end{figure}

\textcolor{black}{The results of the meshless KNN presented through this paper are obtained by enforcing zero divergence as constraint and penalty. Imposing penalties is computationally inexpensive, thus the condition is promoted on all particle locations. On the other hand, a trade-off between accuracy and computational cost must be sought when setting constraints. For this reason, throughout the work, the zero divergence condition is set as a constraint only on $10\%$ of the original particle distribution ($\approx 100$).  }

\begin{figure}[t]
\centering
\begin{overpic}[width=0.95\textwidth, unit = 1mm]{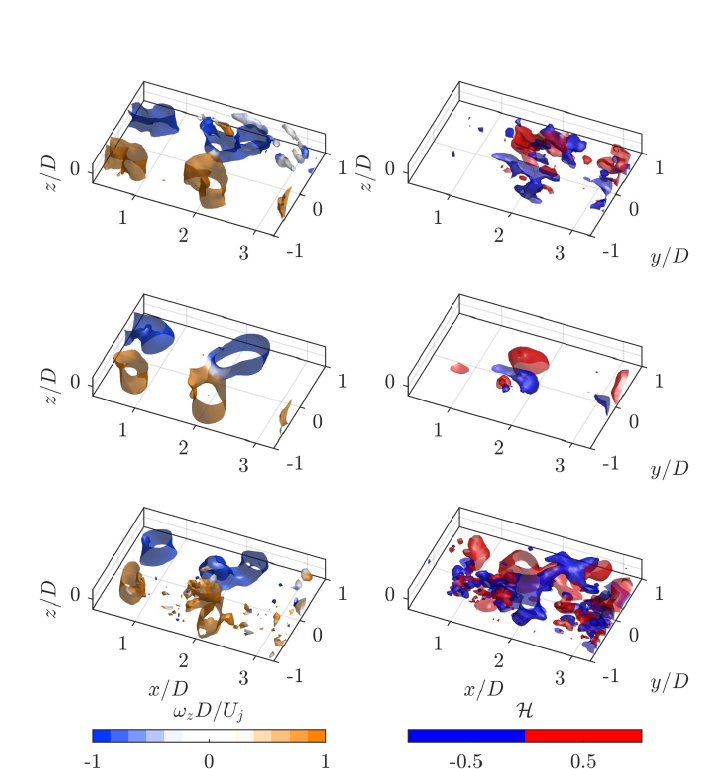}
	\put(-10,123){\parbox{50mm}{  \flushleft \textnormal{\textbf{ a) PIV IW = 64}}}}
	\put(-10,85){\parbox{50mm}{  \flushleft \textnormal{\textbf{ b) Constrained}}}}
	\put(-10,48){\parbox{50mm}{  \flushleft \textnormal{\textbf{ c) Unconstrained}}}}
	
\end{overpic}
\caption{Q-criterion visualization for positive values of $Q$, coloured by the crosswise vorticity $\omega_zD/U_j$  (first column) and isocontours of positive and negative helicity $\mathcal{H}$  (second column). Depicted cases: a) reference \ac{PIV} with interrogation window of $64$ voxels, b) meshless KNN-PTV with penalties and constraints and c) meshless KNN-PTV without penalties and constraints.   }
\label{fig.QHConstvsUnc}
\end{figure}

\begin{figure}[t]
\centering
\begin{tikzpicture}
	\begin{axis}[
		xlabel={$n_\lambda$},
		ylabel={$\tau$},
		xmin=0, xmax=1000,
		ymin=0.9, ymax=1.5,
		xtick={0,100,...,1000},
		ytick={1,1.1,...,1.5},
		grid=both,
		grid style={line width=.1pt, draw=gray!40},
		major grid style={line width=.2pt, draw=gray!60},
		width=10cm,
		height=6cm,
		tick label style={font=\small},
		label style={font=\small},
		legend style={font=\small, at={(0.97,0.03)}, anchor=south east},
		]
		
		% Data plot
		\addplot[
		color=blue,
		mark=*,
		thick
		] coordinates {
			(0,1.0000)
			(100,1.0487)
			(200,1.1058)
			(300,1.1496)
			(400,1.1935)
			(500,1.2373)
			(600,1.2759)
			(700,1.3144)
			(800,1.3529)
			(900,1.3915)
			(1000,1.4300)
		};

	\end{axis}
\end{tikzpicture}
\caption{Normalized average reconstruction time $\tau$ versus the number of applied constraints $n_\lambda$. }
\label{fig:timevsconst}
\end{figure}

\textcolor{black}{Figure~\ref{fig.ConstvsUnc} presents a comparison of the same snapshot shown in Fig.~\ref{fig.jet} with and without imposing penalties/constraints. The contours of the three velocity components highlight the regularizing effect that physical constraints and penalties play in the regression, leading to a more physically consistent reconstruction. A statistical analysis as in \S~\ref{sub.stat} confirms these findings. Indeed, without applying constraints, it results $\langle\delta_{\text{RMS}}\rangle = 0.1413$,
$\mathrm{Var}(\delta^2) = 0.1033$ and
$\text{max}(\delta_{RMS}) = 1.0442$. This new value of rms suggests a loss in accuracy of $\approx37\%$ when removing penalties and physical constraints. Even more critical than the average error increase, the variance of the squared error grows much more dramatically, more than a threefold increase. This higher variance indicates that the unconstrained formulation produces highly scattered errors across the domain, with local regions where the reconstruction deviates substantially from the reference. This is also confirmed by the higher peak of error. The constrained approach ensures a more uniform and physically-consistent recovery of the jet structures, while the unconstrained case yields inconsistencies, particularly in regions of strong gradients or vortical activity.  In particular, this can also be observed in Fig.~\ref{fig.QHConstvsUnc}, where the effect of the constraints is evaluated on the same snapshot and the same quantities of Fig.~\ref{fig.Q}. While the $Q$ visualization highlights higher noise contamination, the helicity differs substantially from the reference, underlying the propagation of the error on the derivatives, as discussed in \S~\ref{app}. This discussion emphasizes that relying solely on data-driven regression of the particle information can be significantly improved by regularization with physical information. Note that this comparison is carried out on the configuration of the constrained case. While a basis distribution designed specifically for the unconstrained case could potentially mitigate some discrepancies, matching the performance of the constrained case remains improbable. Another aspect to consider is the sparsity of the data. The tests were conducted on the enriched field; therefore, the discrepancies are expected to be more significant on the basic one, which is sparser.} 

\textcolor{black}{As expected, the inclusion of physical constraints comes at the expense of higher computational costs. Figure~\ref{fig:timevsconst} shows the average reconstruction time per snapshot, $\tau$, normalized by the time elapsed in the unconstrained case, as a function of the number of constraints $n_\lambda$ imposed during the regression. The process involves the allocation of basis functions via clustering (Step~3 of the methodology) and the subsequent weighted regression (Step~4). All results were obtained using the same machine and code implementation for both cases. The trend shows an almost linear increase in computational cost. Considering that the average time for the unconstrained case in this configuration is approximately $75$ seconds, this provides an order-of-magnitude reference for the overall computational effort. For this dataset, increasing the number of constraints beyond $100$ does not justify the additional computational effort; therefore, this value has been chosen as a suitable trade-off. It should also be noted that here only the zero-divergence constraint was imposed, which is not expected to strongly affect the $l_2$ norm of the error.} %These results, however, are not easily generalizable: when other constraints such as Dirichlet or Neumann conditions are employed, their impact is expected to be much more pronounced on both accuracy and computational costs.}

\section{Conclusions}

A novel meshless super-resolution technique has been introduced for image velocimetry, combining the strengths of KNN-PTV and constrained \acp{RBF}. Although presented in the context of particle image velocimetry, the technique is of general application to all cases in which data are sampled at scattered locations at different time instants (e.g. with random moving or on/off sensors). This method increases the density of scattered data by "borrowing" particles (or more generally, samples) from similar snapshots, even without time resolution, while strengthening regression robustness by enforcing physical constraints—all within a fully meshless framework. The results show promising improvements in both reconstruction accuracy and spatial resolution, validated through benchmark tests on experimental $3D$ measurements of a jet flow in air.

The main novelty of this approach is mesh independence, offering analytical representations of flow fields that can be easily interpolated and differentiated on any grid, enabling the extraction of high-resolution instantaneous fields and turbulence statistics. The technique is particularly advantageous for $3D$ flow analysis, where the demand for spatial resolution at reasonable computational costs is more critical compared to planar applications. Moreover, the use of constrained regression allows for handling larger interparticle spacing by enforcing flow-physics-based constraints.

The different analyses carried out in the paper demonstrate the following: first, the crucial role of physical constraints in regularizing the reconstructed flow field to address the larger interparticle spacing typical of $3D$ scenarios; second, the artificial increase of particle density by borrowing particles from similar snapshots helps in the reconstruction of the smallest scales, especially in sparse cases (although its performance is depending on the regularization in post-processing); last, the superiority of regression-based techniques (whether weighted or not) over simple moving averages (weighted or not).

It should also be noted that, while the methodologies examined may have reached their peak performance, the results of the fully meshless algorithm hold potential for further enhancement. Future studies will likely focus on identifying the optimal basis for flow field approximation, opening new avenues for improvement. Pressure estimation and the recovery of time resolution from data scattered in space and time are among the most straightforward applications of the method outside the primary scope of achieving high-resolution flow fields.

\section*{Appendix I. Parameters tuning}
\label{sec:param}

\begin{figure}[t]
\centering
\begin{overpic}[width=\textwidth, unit = 1mm]{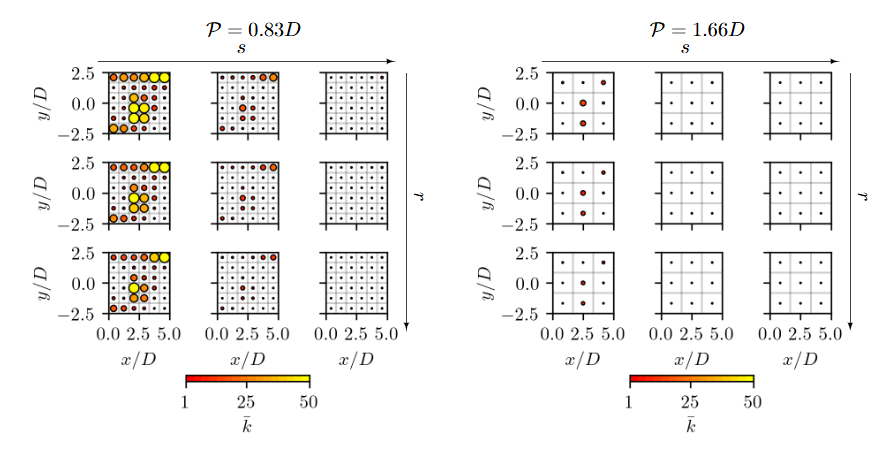}
\end{overpic}
\caption{\textcolor{black}{Mean neighbours value $\bar{k}$ for subdomain size of $0.83$D (left) and $1.66$D (right). The value of the similarity threshold $s$ increases from left to right ($0.6$, $0.75$ and $0.9$). The value of the energy content for truncation increases from top to bottom ($0.85$, $ 0.9$ and $0.95$).} }
\label{fig.param}
\end{figure}

\textcolor{black}{
This appendix aims to evaluate the influence of three parameters involved in the algorithm: the dimension of the reduced feature space $r$, the subdomain size $\mathcal{P}$ and the similarity threshold $s$.
\\
To evaluate the effects of these parameters, a total of $27$ combinations are tested varying the following values: $\mathcal{P} = [0.83D, 1.66D, 5D]$, $r = [0.85, 0.9, 0.95]$, and $s = [0.6, 0.75, 0.9]$.
Regarding $\mathcal{P}$, the smallest value corresponds to partitioning the domain into $6 \times 6$ equal volumes, the intermediate value corresponds to a $3 \times 3$ division, and the largest represents the entire domain without subdivision. Smaller subdomains are not reported, as the reduced number of particles within the volume risks limiting reliable feature extraction, and the restricted spatial extent tends also to exclude larger flow structures that are crucial for accurate reconstruction. In any case, this is the parameter that mostly affects the computational costs of the algorithm, because it implies repeating the feature extraction for each volume.
\\
The selection of $r$ was guided by classical rules of thumb from POD decomposition, considering that the feature set is built upon its modes. Consequently, we chose values based on the cumulative energy captured by the retained modes. Values that are too low result in an overly compressed feature set and significant information loss, whereas those that are too high decrease the effectiveness of dimensionality reduction by retaining less pertinent modes.
\\
Lastly, the values chosen for the similarity threshold categorize cases into low, medium, and high similarity. Setting a low threshold, of course, would lead to aggregate incoherent subvolumes. On the other hand, demanding high similarity would hinder the full exploitation of the potential benefits of the method, especially in datasets of reduced size.
\\
Figure~\ref{fig.param} summarizes the effects of the different parameter configurations.
Specifically, the plots report the average number of neighbors $\bar{k}$ that satisfy the similarity criteria for each subdomain. For clarity, note that $\bar{k} = 1$ implies that, on average, a subdomain only finds itself as a suitable neighbor.
\\
For brevity, the case without domain partitioning is not shown, as it consistently results in $\bar{k} = 1$ due to the stricter similarity constraint applied to the entire domain.
\\
The plots are organized such that the similarity threshold increases from left to right, while the rank of the feature set increases from top to bottom. The background grid indicates the dimensionless size of each subdomain, offering a visual reference of its proportion relative to the full domain.
\\
The most evident trend from this comparison is that increasing the subdomain size reduces the availability of suitable neighbors, necessitating a relaxation of the similarity threshold. This behavior reflects the increasing difficulty of exploiting similarity across larger scales. However, this observation should not be misinterpreted as a motivation to minimize the subdomain size indefinitely. While smaller subdomains indeed facilitate the identification of local similarities, reducing the subdomain size at a fixed particle density leads to fewer particles per volume. Particle availability is crucial for both the similarity assessment and the meshless-POD-based training, as both rely on sufficient sampling to achieve convergence. This challenge is conceptually analogous to the trade-offs encountered when binning flow fields: smaller bins help resolve finer structures, but require enough samples to yield statistically meaningful estimates.
\\
Moreover, $\mathcal{P}$ is the parameter that most significantly impacts the computational cost. Smaller subdomains imply a greater number of regions to process, each requiring local regression and its own meshless POD decomposition. Hence, the final choice must balance the accuracy of the reconstruction with the associated computational burden.
In this study, we selected a subdomain size of $0.83D$, which provides an average of approximately $30$ particles per volume. This configuration ensures a robust similarity evaluation while keeping an acceptable computational costs. Future research may explore adaptive partitioning strategies to better balance local similarity, accuracy and computational efficiency.
\\
Focusing on the case with the smallest subdomains, the influence of the similarity threshold becomes particularly evident: as expected, increasing this threshold (i.e., moving from left to right in the plots) leads to a sharp reduction of  $\bar{k}$, which can ultimately converge to the case where no neighbors are found. This occurs because it becomes increasingly difficult for the algorithm to identify subdomains that exceed the similarity requirement.
\\
There are three main ways to address this issue: lowering the similarity threshold, reducing the rank of the feature space, or adjusting the subdomain size. The rationale behind the first and second options is relatively straightforward. Relaxing the threshold directly increases the likelihood of identifying neighbors, while reducing the dimensionality of the feature space simplifies the similarity search by focusing on dominant characteristics. As discussed earlier, shrinking the subdomain size can help in identifying local similarities but comes with limitations in terms of particle availability and computational cost.
\\
Interestingly enough, while reducing $r$ and $s$ aids neighbor detection, excessive compression or relaxation on the similarity compromises the distinctive feature of the algorithm, i.e., merging \textit{coherent} information. For this reason, configurations with the highest number of neighbors are not necessarily those yielding the best performance, a nuance not reported here for brevity.
\\
Among the three parameters, the similarity threshold $s$ has the most significant impact, both on the reconstruction quality and on the computational costs. In this study, we adopt a threshold of $s = 0.75$, which represents a good compromise for $3D$ datasets. For planar applications, higher thresholds may be feasible. Regarding the feature rank $r$, its influence was found to be negligible in the canonical ranges, and we therefore adopt a conventional choice of $90\%$ cumulative energy retention.
}
\textcolor{black}{\\In summary, while an initial visual inspection combined with physical knowledge, neighbors analysis (map and snapshots), and corresponding enriched field can provide essential hints, fine-tuning this parameter offers diminishing returns once a reasonable result has been achieved, especially considering the additional computational costs. The method is quite robust against minor variations around the optimal or even suboptimal thresholds, thanks to the introduction of the weighting scheme that tends to eliminate the influence of the farthest neighbours. On the other hand, if the user prefers a less empirical approach to extract the best possible results, the same uncertainty-based criterion proposed in \citet{tirelli2023end} could be implemented. That work demonstrated that the uncertainty closely follows the error distribution, meaning that the ``elbow'' in the uncertainty curve is a good indicator of the optimal parameter value. In the present implementation, however, we opted not to rely on this criterion because it is computationally expensive: it requires multiple test datasets to evaluate different conditions and identify the value of the parameter that minimizes the uncertainty.  Instead, leveraging a combination of user expertise and visually guided parameter selection (as common in PIV processing), following the provided rules-of-thumb, allows a dramatic reduction in the computational cost without significantly degrading the quality of the reconstructions.}

\section*{Appendix II. Radial basis functions derivatives}
\textcolor{black}{
PIV velocity measurement carries an intrinsic uncertainty, amplified when computing differential quantities using local finite differences. This happens because such schemes estimate derivatives by subtracting neighboring velocity values, both of which are noisy, and dividing by a small spatial increment. Consequently, the uncertainty scales inversely with the local spacing, so increasing spatial resolution inherently amplifies noise in the computed gradients.
}

\textcolor{black}{
To mitigate this limitation, one could increase the particle density and reduce the interrogation window size. However, this introduces correlation between neighboring velocity vectors from overlapping particle images, leading to biased estimates in regions with strong gradients or sparse seeding. Such biases propagate to derivative computations, degrading differential fields \citep[Chap.~6.4]{raffel2018particle}.
}

\textcolor{black}{
The availability of an analytical approximation of the flow field can be exploited to compute its spatial derivatives and estimate derived quantities. Once the analytical approximation of the fields are obtained, their derivatives can be readily calculated from the derivatives of the basis functions in Eq.~\ref{eq.gaussian}:
}

\begin{subequations}
\label{eq:gaussian_derivatives}
\begin{align}
\frac{\partial}{\partial x} \varphi_j(\mathbf{x}; \mathbf{X}_{c,j}^{(k)}) 
&= -2 c^2 (x - X_{c,j,1}^{(k)}) \, \varphi_j(\mathbf{x}; \mathbf{X}_{c,j}^{(k)}) \label{eq:gaussian_dx} \\
\frac{\partial}{\partial y} \varphi_j(\mathbf{x}; \mathbf{X}_{c,j}^{(k)}) 
&= -2 c^2 (y - X_{c,j,2}^{(k)}) \, \varphi_j(\mathbf{x}; \mathbf{X}_{c,j}^{(k)}) \label{eq:gaussian_dy}\,\,. \\
\frac{\partial}{\partial z} \varphi_j(\mathbf{x}; \mathbf{X}_{c,j}^{(k)}) 
&= -2 c^2 (z - X_{c,j,3}^{(k)}) \, \varphi_j(\mathbf{x}; \mathbf{X}_{c,j}^{(k)}) \label{eq:gaussian_dz}
\end{align}
\end{subequations}

\textcolor{black}{
This leads to the following general expression for the partial derivative of a velocity component along the spatial coordinates:
}

\begin{equation}
\frac{\partial u_\alpha}{\partial x_\beta}(\mathbf{X}^{(k)}, \bm{t}_k) = 
\sum_{j=1}^{N_b^{(k)}} \,  -2 c^2 (x_\beta - X_{c,j,\beta}^{(k)}) \, \varphi_j(\mathbf{x}; \mathbf{X}_{c,j}^{(k)})a_{j,\alpha}(\bm{t}_k), 
\label{eq:velocity_partial_derivative}
\end{equation}

\textcolor{black}{
\noindent where $ \alpha, \beta \in \{1,2,3\} $ denote the velocity component and the spatial direction, respectively. This formulation enables the analytical approximation of all quantities related to the spatial derivatives of the velocity field. For instance, quantities involving both velocity and its derivatives, such as pressure, can be computed following the approach described in \citet{sperotto2022meshless}.
}

\section*{Data Availability Statements}
The data that support the findings of this article are openly available \cite{tirelli_2026_18328319}. The codes are available on GitHub \url{https://github.com/erc-nextflow/MESHLESS-KNN-PTV}.

\section*{Declaration of AI use}
During the preparation of this work, the authors used Chat-GPT and Grammarly to improve the readability and language of this manuscript. After using this tool/service, the authors reviewed and edited the content as needed and took full responsibility for the publication’s content.

\section*{Acknowledgment}
This project has received funding from the European Research Council (ERC) under the European Union’s Horizon 2020 research and innovation program (grant agreement No 949085). Views and opinions expressed are however those of the authors only and do not necessarily reflect those of the European Union or the European Research Council. Neither the European Union nor the granting authority can be held responsible for them.

\bibliography{sn-bibliography}% common bib file

\end{document}